\documentclass[12pt,a4paper]{article}
\usepackage[utf8]{inputenc}
\usepackage{graphicx}
\usepackage{epstopdf}
\usepackage{hyperref}
\usepackage{multirow}
\usepackage{geometry}
\usepackage{color}

\usepackage{amsmath}
\allowdisplaybreaks[1]

\begin{document}

\begin{titlepage}
\begin{flushright}
LU TP 17-04\\
28th September 2017
\end{flushright}
\vfill
\begin{center}
{\Large\bf An Analytic Analysis of the Pion Decay Constant\\[3mm]in Three-Flavoured Chiral Perturbation Theory}

\vfill
{\bf B. Ananthanarayan$^a$, Johan Bijnens$^b$ and Shayan Ghosh$^a$}\\[1cm]
{$^a$ Centre for High Energy Physics, Indian Institute of Science,\\Bangalore-560012, Karnataka, India}\\[0.5cm]
{$^b$Department of Astronomy and Theoretical Physics,\\Lund University,
S\"olvegatan 14A, SE 223-62 Lund, Sweden}
\end{center}
\vfill

\begin{abstract}
A representation of the two-loop contribution to the pion decay constant in $SU(3)$ chiral perturbation theory is presented. The result is analytic upto the contribution of the three (different) mass sunset integrals, for which an expansion in their external momentum has been taken. We also give an analytic expression for the two-loop contribution to the pion mass based on a renormalized representation and in terms of the physical eta mass. We find an expansion of $F_{\pi}$ and $M_{\pi}^2$ in the strange quark mass in the isospin limit, and perform the matching of the chiral SU(2) and SU(3) low energy constants. A numerical analysis demonstrates the high accuracy of our representation, and the strong dependence of the pion decay constant upon the values of the low energy constants, especially in the chiral limit. Finally, we present a simplified representation that is particularly suitable for fitting with available lattice data.
\end{abstract}
\vfill
\vfill
\end{titlepage}

\section{Introduction}

The mass and decay constants of the pions, kaons and the eta have been worked out to two-loop accuracy in three-flavoured chiral perturbation theory (ChPT) in \cite{Amoros:1999dp} some time ago. The expressions for these at this order bring in a class of diagrams known as the sunsets. For the decay constants, in addition to the sunset integral, derivatives of the sunsets with respect to the square of the external momentum (also known as `butterfly' diagrams), evaluated at a value equal to the square of the mass of the particle in question, are needed. The sunset diagrams themselves have been studied in field theory literature for many years now, and for particular mass configurations analytic expressions exist in Laurent series expansions in $\epsilon = (4-d)/2$. In general, however, the sunsets and their derivatives have to be evaluated numerically and a publicly available software \cite{Bijnens:2014gsa} does this with user driven inputs.

There is, however, a need for an analytic study of the observables in ChPT since one would like to have an intuitive sense for the results appearing therein. More importantly, with recent advances allowing lattice simulations to tune the quark masses to near physical values, a combining of lattice and ChPT results has become possible. However, at next to next to leading order (NNLO), three flavoured ChPT amplitudes are available only numerically or take a complicated form, and thus have not been used much by the lattice community. With this in mind, \cite{Ecker:2010nc,Ecker:2013pba} has advocated a large $N_c$ motivated approach to replace the two-loop integrals by effective one-loop integrals, and find it fruitful for the study of the ratio $F_K/F_\pi$ as well as $F_\pi$. The analytic studies of SU(3) amplitudes in the strange quark mass expansion of \cite{Gasser:2007sg, Gasser:2009hr, Gasser:2010zz} are also steps in that direction, but as the results presented there are in the chiral limit $m_u=m_d=0$, the need for more general expressions is left unfulfilled.

Some years ago, Kaiser \cite{Kaiser:2007kf} studied the problem of the pion mass in the analytic framework, and was able to employ well known properties of sunset integrals to reduce a large number of expressions to analytic ones. One exception was the sunset integral with kaons and an eta propagating in the loops with the external momentum at $s=m_\pi^2$, for which an expansion around $m_\pi^2$ was used. Kaiser \cite{Kaiser:2007kf} also replaced the $m_\eta$ in his work by the leading order Gell-Mann-Okubo (GMO) formula. In principle, therefore, one can get an expansion in $m_{\pi}^2$ to arbitrary accuracy, proving thereby the accessibility of an analytical approach to the full two-loop result. For practical purposes, we have used the expansion up to and including $m_\pi^4$ terms. These are more than sufficient for the numerical accuracy wanted.

The reason why it is possible to attain the objectives above is that for many purposes, the sunset integrals are accessible analytically for kinematic configurations known as threshold and pseudo-threshold configurations \cite{Berends:1997vk}, as well as for the case when the square of the external momentum vanishes \cite{Davydychev:1992mt}. Indeed, this is the case for most of the sunset integrals appearing in the expressions for the mass and decay constants. These properties also allow one to isolate the divergent parts in closed form, while the finite part remains calculable in analytic form only for special cases.  On the other hand, there is always an integral representation for the finite part which can be evaluated numerically. Furthermore, for the most general case, all sunsets can be reduced to a set of master integrals.  All other vector and tensor integrals, as well their derivatives with respect to the square of the external momentum, can also be reduced to master integrals. The work of \cite{Tarasov:1997kx} in developing this work is noteworthy, as is the automation of these relations with the publicly available Mathematica package Tarcer \cite{Mertig:1998vk}. Application of these methods and tools to sunset diagrams in chiral perturbation theory is elucidated in \cite{Ananthanarayan:2016pos}.

Inspired by the developments above, we now seek to extend the work of \cite{Kaiser:2007kf} for the case of the pion decay constant in an expansion around $s=0$, which also brings in the butterfly diagrams.  In contrast to the approach of \cite{Kaiser:2007kf}, we will retain the mass of the eta without recourse to the GMO. This is the main objective of the present work. As a side result, we also give the expression for the two-loop pion mass with the full eta mass dependence.

In principle, this may be also extended to the mass and decay constant of the kaon and the eta, but the expansion about $s=0$ for these particles when particles of unequal mass are running around in the loops is bound to converge poorly, and one would have to go to very high orders in the expansion, thereby losing the appeal of such a result. Thus we confine ourselves to the pion in this work. We present expressions for the kaon and eta masses and decay constants in a future publication \cite{ABFG:2017}.

As an application of the expressions given here, we give their expansion in the strange quark mass in the isospin limit and perform the `matching' of the three flavoured low energy constants $F_0$ and $B_0$ with their two flavoured counterparts $F$ and $B$, respectively. We compare our results with those given in \cite{Kaiser:2006uv} and the chiral limit results of \cite{Gasser:2007sg}. The results given in this work, however, go beyond the chiral limit matching done in the aforementioned papers. Indeed, the full expressions presented here allow for an expansion up to an arbitrary order in the quark masses.

The scheme of this paper is as follows. In Section~\ref{SecSunsets} we briefly review sunset diagrams and their evaluation. In Section~\ref{SecPionDecay} we give the expressions for the analytical results up to $\mathcal{O}(m_{\pi}^4)$ for the pion decay constant at two loops. We repeat the analysis for the two-loop pion mass contribution in Section~\ref{SecPionMass}. In Section~\ref{SecExp}, we give the s-quark expansion for both the pion decay constant as well as the pion mass, and perform the matching of the two- and three- flavour low-energy constants (SU(2) and SU(3) LECs). We present a numerical analysis of our results in Section~\ref{SecNum}, and in Section~\ref{SecLatticeFit} we discuss the fitting of lattice data with the expressions given in this paper, and present them in a form that allows one to perform these fits relatively easily. In Section~\ref{SecLatticeFit}, we discuss several possible ways of expressing the results of this paper, and present a simplified representation that is particularly suitable for performing fittings with available lattice data. We conclude in Section~\ref{SecConc} with a discussion of possible future work in this area.

\section{Sunset Diagrams and their Derivatives \label{SecSunsets}}

\begin{figure}[tb]
\centering
\includegraphics{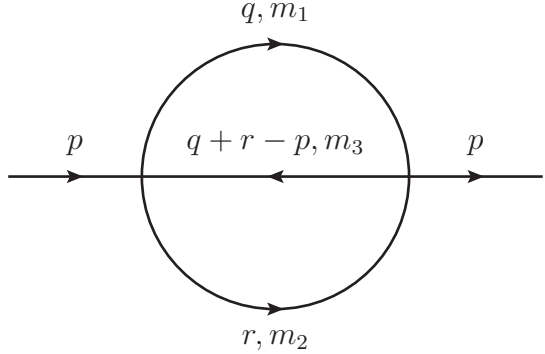} 
\caption{The two-loop self energy ``sunset'' diagram}
\label{FigSunset}
\end{figure}

The sunset diagram, shown in Figure~\ref{FigSunset}, represents the two-loop Feynman integral:
\begin{align}
H_{\{\alpha,\beta,\gamma\}}^d (m_1,m_2,m_3;s) = \frac{1}{i^2} \int \frac{d^dq}{(2\pi)^d} \frac{d^dr}{(2\pi)^d} \frac{1}{[q^2-m_1^2]^{\alpha} [r^2-m_2^2]^{\beta} [(q+r-p)^2-m_3^2]^{\gamma}} 
\label{sunsetdef}
\end{align}

Aside from the basic scalar integral, there exist tensor varieties of the sunset integral with loop-momenta in the numerator. The two tensor integrals that are of relevance to this work are $H_{\mu}$ and $H_{\mu\nu}$, in which the momenta $q_{\mu}$ and $q_{\mu} q_{\nu}$, respectively, appear in the numerator. These may be decomposed into linear combinations of scalar integrals via the Passarino-Veltman decomposition as:
\begin{align}
	& H_{\mu}^d = p_{\mu} H_1 \nonumber \\
	& H_{\mu\nu}^d = p_{\mu}p_{\nu} H_{21} + g_{\mu\nu} H_{22} \label{H21}
\end{align}

The representation of the pion decay constants in \cite{Amoros:1999dp} involves the scalar integrals $H_1$ and $H_{21}$. Taking the scalar product of $H_{\mu}^d$ with $p^{\mu}$ allows us to express the integral $H_1$ in terms of the sunset integral with the scalar numerator $q.p$. Similarly, we may express $H_{21}$ in terms of sunset integrals with numerators $(q.p)^2$ and $q^2$:
\begin{align}
	& H_1 = \frac{\langle \langle q.p \rangle \rangle}{p^2} \nonumber \\
	& H_{21} = \frac{\langle \langle (q.p)^2 \rangle \rangle d - \langle \langle q^2 \rangle \rangle p^2}{p^4 (d-1)} 
\end{align}
where $\langle \langle X \rangle \rangle$ represents a sunset integral with numerator $X$.

Another class of integrals that appear in the representation of \cite{Amoros:1999dp} is the derivative of the sunset integrals and the $H_1$ and $H_{21}$ with respect to the external momentum. In some places in the literature, these are sometimes known as `butterfly' diagrams. These butterfly integrals may be expressed as sunset integrals of higher dimension by means of the following expression, which can be derived from the Feynman parameter representation of the sunset integrals, and a more general version of which is given in \cite{Kaiser:2007kf}.
\begin{align}
	\left( \frac{\partial}{\partial s} \right)^n H_{\{\alpha,\beta,\gamma\}}^d = (-1)^n (4 \pi)^{2n} \frac{\Gamma(\alpha+n) \Gamma(\beta+n) \Gamma(\gamma+n)}{\Gamma(\alpha) \Gamma(\beta)\Gamma(\gamma)} H_{\{\alpha+n,\beta+n,\gamma+n\}}^{d+2n}
\end{align}

Tarasov \cite{Tarasov:1997kx} has shown that by means of integration by parts relations, all sunset integrals may be expressed as linear combinations of four master integrals, namely $H_{\{1,1,1\}}^d$, $H_{\{2,1,1\}}^d$, $H_{\{1,2,1\}}^d$ and $H_{\{1,1,2\}}^d$, and the one-loop tadpole integral:
\begin{align}
	A^d(m) = \frac{1}{i} \int \frac{d^d q}{(2\pi)^d} \frac{1}{q^2 - m^2} = - \frac{ \Gamma \left( 1-d/2 \right)}{(4\pi)^{d/2}} m^{d-2}
\end{align}

This includes sunset integrals of dimensions greater than $d$, permitting us to express the butterfly integrals in terms of the four master integrals and tadpoles. Scalar sunset integrals with non-unit numerators, such as those appearing in Eq.(\ref{H21}) may also be expressed in terms of the four master integrals and tadpoles. The Tarcer package \cite{Mertig:1998vk}, written in Mathematica, automates the application of Tarasov's relations, and we have made extensive use of it in this work. We have also made use of the package Ambre \cite{Gluza:2007rt, Gluza:2010rn}, which allows for a direct evaluation of many scalar and tensor Feynman integrals using a Mellin-Barnes approach, to numerically check our breakdown of the sunset and butterfly diagrams into master integrals. The theory of analytic (rather than numeric) evaluation of multi-fold Mellin-Barnes integrals is described with examples in \cite{Friot:2011ic, Aguilar:2008qj}.

As is the usual practice in chiral perturbation theory, we use a modified version of the $\overline{MS}$ scheme to handle the divergences arising from the evaluation of the sunset diagrams. The subtraction procedure to two-loop order in ChPT is equivalent to multiplying Eq.(\ref{sunsetdef}) by $(\mu_{\chi}^2)^{4-d}$, where:
\begin{align}
	\mu_{\chi}^2 \equiv \mu^2 \frac{e^{\gamma_E - 1}}{4\pi}
\end{align}
and taking into consideration only the $\mathcal{O}(\epsilon^0)$ part of the result in a Laurent expansion about $\epsilon = 0$. We denote such renormalized sunset integrals by use of the subscript $\chi$ instead of $d$, i.e.
\begin{align}
	H^{\chi}_{\{a,b,c\}} \equiv (\mu_{\chi}^2)^{4-d} H^d_{\{a,b,c\}}
\end{align}

The inclusion of factor $\mu$ raised to a power of the dimension $d$ introduces terms involving chiral logarithms, i.e.
\begin{align}
	l_P^r \equiv \frac{1}{2(4\pi)^2} \log \left[ \frac{m_P^2}{\mu^2} \right] \qquad P = \pi, K, \eta
	\label{chiralLog}
\end{align}
In the results presented in this paper, we group together all terms containing chiral logarithms, whether or not they arise from the renormalized sunset integrals. We therefore use the notation:
\begin{align}
	H^{\chi}_{\{a,b,c\}} \equiv \overline{H}^{\chi}_{\{a,b,c\}} + H^{\chi,\log}_{\{a,b,c\}} \label{Hnotation}
\end{align}
where $H^{\chi,\log}$ are the terms of the sunset integral containing chiral logarithms, and $\overline{H}^{\chi}$ is the aggregation of the remainder. All results given hereafter have been renormalized using this subtraction scheme, and are presented using the notation above.

Analytic expressions for the master integrals themselves have been studied thoroughly, and several results exist in the literature \cite{Berends:1997vk, Davydychev:1992mt, Gasser:1998qt, Czyz:2002re, Martin:2003qz, Adams:2015gva}. For sunset integrals with only one mass scale, there is a further reduction in the number of master integrals, and all sunsets can be expressed in terms of the tadpole integral, $A^{\chi} = \mu_{\chi}^{4-d} A^d$, and $H^{\chi}_{\{1,1,1\}}$, which is given in \cite{Berends:1997vk, Gasser:1998qt}, amongst others, as:
\begin{align}
	H^{\chi}_{\{1,1,1\}} = - \left( \mu^2 e^{\gamma_E-1} \right)^{2\epsilon} \frac{(m^2)^{1-2\epsilon}}{(4\pi)^4} \frac{\Gamma^2(1+\epsilon)}{(1-\epsilon)(1-2\epsilon)} \bigg( - \frac{3}{2\epsilon^2} + \frac{1}{4\epsilon} + \frac{19}{8} \bigg) + \mathcal{O}(\epsilon)
\end{align}
Analytic expressions for the two mass scale integrals can be found by means of the pseudothreshold results of \cite{Berends:1997vk}.

Expressions for the three mass sunset integrals are given in \cite{Adams:2015gva} in terms of elliptic dilogarithmic functions. However, as one of the principal reasons for the lack of use of ChPT results by the lattice community is the complicated form of many of the results, we wish to keep the expression derived here as simple and accessible as possible. To this end, and to stay true to the spirit of the method of \cite{Kaiser:2007kf}, instead of using the results of \cite{Adams:2015gva} we take an expansion in the external momentum $s$ upto order $\mathcal{O}(s^2)$:
\begin{align}
	H^{\chi}_{\{\alpha,\beta,\gamma\}} = K_{\{\alpha,\beta,\gamma\}} + s \, K'_{\{\alpha,\beta,\gamma\}} + \frac{s^2}{2!} \, K''_{\{\alpha,\beta,\gamma\}} + \mathcal{O}(s^3)  
\end{align}
where $K_{\{\alpha,\beta,\gamma\}} \equiv H^{\chi}_{\{\alpha,\beta,\gamma\}} |_{s=0}$. In this special case of $s=0$, as in the case of the single mass scale sunsets, all sunset integrals may be expressed solely in terms of $K_{\{1,1,1\}}$ and tadpole integrals \cite{Tarasov:1997kx}.

The pion mass and decay constant at two loops both involve a sunset integral with the following three mass scale configuration: \[ H^{\chi}_{\{\alpha,\beta,\gamma\}} \left( m_{K}, m_{K}, m_{\eta}; s=m_{\pi}^2 \right) \] This may be expanded in $s$ by making use of the result \cite{Amoros:1999dp, Kaiser:2007kf, Davydychev:1992mt}:
\begin{align}
	\frac{2 \left( 4 \pi \right)^4}{M^2} & H^{\chi}_{\{1,1,1\}}\{M,M,m;0\} \nonumber \\	
	& = \left( 2 + \frac{m^2}{M^2} \right) \frac{1}{\epsilon^2} + \left( \frac{m^2}{M^2} \left( 1 - 2  \log \left[ \frac{m^2}{\mu^2} \right] \right) + 2 \left( 1 - 2  \log \left[ \frac{M^2}{\mu^2} \right] \right) \right) \frac{1}{\epsilon} \nonumber \\
	& - \frac{2}{(\mu^2)^{2\epsilon}} \bigg( \frac{m^2}{M^2} \log \left[ \frac{m^2}{\mu^2} \right] \left( 1 - \log \left[ \frac{m^2}{\mu^2} \right] \right) + 2 \log \left[ \frac{M^2}{\mu^2} \right] \left( 1 - \log \left[ \frac{M^2}{\mu^2} \right] \right) \bigg) \nonumber \\
	& - \frac{m^2}{M^2} \log^2 \left[ \frac{m^2}{M^2} \right] + \left( \frac{m^2}{M^2} - 4 \right) F \left[ \frac{m^2}{M^2} \right] + \left( 2 + \frac{m^2}{M^2} \right) \left( \frac{\pi^2}{6} + 3 \right) + \mathcal{O}(\epsilon) \label{K111}
\end{align}
where 
\begin{align}
F[x] = \frac{1}{\sigma} \bigg[ 4 \text{Li}_2 \bigg( \frac{\sigma-1}{\sigma+1} \bigg) + \log^2 \bigg( \frac{1-\sigma}{1+\sigma} \bigg) + \frac{\pi^2}{3} \bigg] , \qquad \sigma = \sqrt{1-\frac{4}{x}}  \label{Fdef}
\end{align}

\section{The Pion Decay Constant to Two Loops \label{SecPionDecay}}

The pion decay constant is given in \cite{Amoros:1999dp} as:
\begin{align}
	F_{\pi}=F_0(1+\overline{F}_{\pi}^{(4)}+\overline{F}_{\pi}^{(6)}) + \mathcal{O}(p^8)
\end{align}
where the $\mathcal{O}(p^6)$ contribution can be broken up into a piece that results from the model-dependent counterterms $(\overline{F}_{\pi}^{(6)})_{CT}$, and one that results from the chiral loop $(\overline{F}_{\pi}^{(6)})_{loop}$. For the pion, the explicit form of these terms are given by:
\begin{align}
F_{\pi}^2 \overline{F}_{\pi}^{(4)} = 4 m_{\pi}^2 (L^r_{4}+L^r_{5})+8 L^r_{4} m_{K}^2-l^r_{K} m_{K}^2-2 l^r_{\pi} m_{\pi}^2
\end{align}

\begin{align}
	F_{\pi}^4 (\overline{F}_{\pi})^{(6)}_{CT} &= 8 m_{\pi}^4 (C^r_{14}+C^r_{15}+3 C^r_{16}+C^r_{17})+16 m_{K}^2 m_{\pi}^2 (C^r_{15}-2 C^r_{16})+32 C^r_{16} m_{K}^4
\end{align}
where $m_{P}$ with $P=\pi,K,\eta$ are the physical meson masses, and $l_{P}^{r}$ are the chiral logarithms defined in Eq.(\ref{chiralLog}). Note that the $C_i$ used in this paper are dimensionless.

The loop contributions can be subdivided as follows: 
\begin{align}
	F_{\pi}^4 (\overline{F}_{\pi})^{(6)}_{loop} = \overline{d}^{\pi}_{sunset} + d^{\pi}_{log \times log} + d^{\pi}_{log} + d^{\pi}_{log \times L_i} + d^{\pi}_{L_i} + d^{\pi}_{L_i \times L_j} \label{dloop}
\end{align}

The terms containing the LECs $L_i$ but no chiral logarithms are given by:
\begin{align}
	(16 \pi^2) d^{\pi}_{L_i} = \frac{8}{9} \left(L^r_{2}+\frac{L^r_{3}}{3}\right) m_{K}^2 m_{\pi}^2 - \left(2 L^r_{1}+\frac{37}{9} L^r_{2} + \frac{28}{27} L^r_{3} \right) m_{\pi}^4 - \left( \frac{52}{9} L^r_{2}+\frac{43}{27} L^r_{3} \right) m_{K}^4
\end{align}
and the terms bilinear in the LECs are contained in:
\begin{align}
	d^{\pi}_{L_i \times L_j} &= 32 m_{K}^2 m_{\pi}^2 \left(7 (L^r_{4})^2+5 L^r_{4} L^r_{5}-8 L^r_{4} L^r_{6}-4 L^r_{5} L^r_{6}\right)+32 m_{K}^4 L^r_{4} (7 L^r_{4}+2 L^r_{5}-8 L^r_{6}-4 L^r_{8}) \nonumber \\
	& \quad +8 m_{\pi}^4 (L^r_{4}+L^r_{5}) (7 L^r_{4}+7 L^r_{5}-8 L^r_{6}-8 L^r_{8})
\end{align}

The remaining three terms of Eq.(\ref{dloop}) give the terms containing the chiral logs. Explicitly, the following gives the terms linear in chiral logarithms:
\begin{align}
	(16 \pi^2) d^{\pi}_{log} &= m_{K}^4 \left(\frac{2}{3} l^r_{\eta} + \frac{23}{8} l^r_{K} + \frac{9}{8} l^r_{\pi} \right)+ m_{K}^2 m_{\pi}^2 \left(\frac{139}{72} l^r_{\pi} - \frac{1}{72} l^r_{\eta} - \frac{1}{2} l^r_{K} \right) + m_{\pi}^4 \left(\frac{1381}{288} l^r_{\pi} - \frac{11}{288} l^r_{\eta} \right) \label{dlogGMO}
\end{align}
while the terms bilinear in the $l_P^r$ are contained in:
\begin{align}
	d^{\pi}_{log \times log} &= m_{K}^4 \left(\frac{7}{72} (l^r_{\eta})^2 - \frac{55}{36} l^r_{\eta} l^r_{K} + \frac{5}{36} (l^r_{K})^2 - \frac{3}{4} l^r_{K} l^r_{\pi} + \frac{3}{8} (l^r_{\pi})^2 \right) + m_{\pi}^4 \left(  \frac{41}{8} (l^r_{\pi})^2 - \frac{1}{24}(l^r_{\eta})^2 \right) \nonumber \\
	& + m_{K}^2 m_{\pi}^2 \left(\frac{1}{9}(l^r_{\eta})^2 + \frac{4}{9} l^r_{\eta} l^r_{K} + \frac{1}{9} (l^r_{K})^2 +\frac{25}{3} l^r_{K} l^r_{\pi} - \frac{7}{6} (l^r_{\pi})^2 \right)  + \frac{1}{2} \frac{m_{K}^6}{m_{\pi}^2} \left( l^r_{\eta} - l^r_{K} \right)^2 \label{dloglogGMO}
\end{align}

The contributions from terms involving products of chiral logarithms and the LECs are collected in:
\begin{align}
	d^{\pi}_{log \times L_i} & = 4 m_{\pi}^4 l^r_{\pi} (14 L^r_{1}+8 L^r_{2}+7 L^r_{3}-13 L^r_{4}-10 L^r_{5}) + \frac{4}{9} \left(4 m_{K}^2-m_{\pi}^2\right)^2 l^r_{\eta}  (4 L^r_{1}+L^r_{2}+L^r_{3}-3 L^r_{4}) \nonumber \\
	& + 4 m_{K}^4 l^r_{K} (16 L^r_{1}+4 L^r_{2}+5 L^r_{3}-14 L^r_{4}) - m_{K}^2 m_{\pi}^2 (4 l^r_{K} (3 L^r_{4} + 5 L^r_{5}) + 48 l^r_{\pi} L^r_{4})
\end{align}

And finally, the contributions from the sunset diagrams are given in:
\begin{align}
	{d}^{\pi}_{sunset} & = \frac{1}{(16\pi^2)^2} \bigg( \frac{35}{288} m_{\pi}^4 \pi^2 + \frac{41}{128} m_{\pi}^4 + \frac{1}{144} m_{\pi}^2 m_{K}^2  \pi^2 - \frac{5}{32} m_{\pi}^2 m_{K}^2 + \frac{11}{72} m_{K}^4 \pi^2 + \frac{15}{32} m_{K}^4 \bigg) \nonumber \\
	& + \frac{5}{12} m_{\pi}^4 \overline{H}'^{\chi}_{\pi \pi \pi} - \frac{1}{2} m_{\pi}^2 \overline{H}^{\chi}_{\pi \pi \pi} - \frac{5}{16} m_{\pi}^4 \overline{H}'^{\chi}_{\pi K K} + \frac{1}{16} m_{\pi}^2 \overline{H}^{\chi}_{\pi K K} + \frac{1}{36} m_{\pi}^4 \overline{H}'^{\chi}_{\pi \eta \eta} \nonumber \\
	& + \frac{1}{2} m_{\pi}^2 m_K^2 \overline{H}'^{\chi}_{K \pi K} - \frac{1}{2} m_{K}^2 \overline{H}^{\chi}_{K \pi K} - \frac{5}{12} m_{\pi}^4 H'^{\chi}_{K K \eta} - \frac{1}{16} m_{\pi}^4 \overline{H}'^{\chi}_{\eta K K} + \frac{1}{4} m_{\pi}^2 m_K^2 \overline{H}'^{\chi}_{\eta K K} \nonumber \\
	& + \frac{1}{16} m_{\pi}^2 \overline{H}^{\chi}_{\eta K K} - \frac{1}{4} m_{K}^2 \overline{H}^{\chi}_{\eta K K} + \frac{1}{2} m_{\pi}^4 {\overline{H}'^{\chi}_{1}}_{\pi K K} + m_{\pi}^4 {\overline{H}'^{\chi}_{1}}_{K K \eta} + \frac{3}{2} m_{\pi}^4 {\overline{H}'^{\chi}_{21}}_{\pi \pi \pi} \nonumber \\
	&  - \frac{3}{16} m_{\pi}^4 {\overline{H}'_{21}}^{\chi}_{\pi K K} + \frac{3}{2} m_{\pi}^4 {\overline{H}'_{21}}^{\chi}_{K \pi K} + \frac{9}{16} m_{\pi}^4 {\overline{H}'_{21}}^{\chi}_{\eta K K}
\end{align}
where we use the notation:
\begin{align}
	\overline{H}^{\chi}_{aP bQ cR} = \overline{H}^{\chi}_{\{a,b,c\}} \{ m_P, m_Q, m_R; s = m_{\pi}^2 \}
\end{align}
with $\overline{H}^{\chi}_{\{a,b,c\}}$ as defined in Eq.~(\ref{Hnotation}).
$a,b,c$ will be suppressed if equal to 1. The terms resulting from the sunset integrals which involving chiral logarithms have been included in  ${d}^{\pi}_{log}$ or ${d}^{\pi}_{log \times log}$ as appropriate.

Evaluating the sunset integrals as described in Section (\ref{SecSunsets}), ${d}^{\pi}_{sunset}$ can be re-expressed as:
\begin{align}
	{d}^{\pi}_{sunset} &= \frac{1}{(16 \pi^2)^2} \bigg[ \left(\frac{3445}{1728}+\frac{107 \pi ^2}{864}\right) m_{K}^4+\left(\frac{125}{864}+\frac{17 \pi ^2}{324}\right) m_{K}^2 m_{\pi}^2  - \left(\frac{3}{2}-\frac{\pi ^2}{12}\right) \frac{m_{K}^6}{m_{\pi}^2} \nonumber \\
	& \qquad -\left(\frac{35}{6912}+\frac{13 \pi ^2}{2592}\right) m_{\pi}^4 \bigg] + {d}^{\pi}_{\pi K K} + {d}^{\pi}_{\pi \eta \eta} + {d}^{\pi}_{K K \eta}
\label{dsunsetGMO}
\end{align}
where
\begin{align}
	{d}^{\pi}_{\pi K K} & = - \left(\frac{9}{16} \frac{m_{K}^4}{m_{\pi}^2} + \frac{3}{4} m_{K}^2 + \frac{1}{48} m_{\pi}^2 \right) \overline{H}^{\chi}_{\pi K K} + \left( \frac{3}{4} m_{K}^4 + \frac{1}{6} m_{K}^2 m_{\pi}^2 +\frac{m_{\pi}^4}{12} \right) \overline{H}^{\chi}_{2\pi K K}
\label{dpkk}
\end{align}

\begin{align}
	{d}^{\pi}_{\pi \eta \eta} & = \left( -\frac{1}{36} m_{\pi}^2 \right) \overline{H}^{\chi}_{\pi \eta \eta}+\left( \frac{1}{36} m_{\pi}^4 \right) \overline{H}^{\chi}_{2\pi \eta \eta}
\label{dpee}
\end{align}

\begin{align}
\label{dkkeGMO}
	{d}^{\pi}_{K K \eta} =& \left( \frac{15}{16} \frac{m_{K}^4}{m_{\pi}^2} - \frac{13}{36} m_{K}^2 + \frac{13}{144} m_{\pi}^2 \right) \overline{H}^{\chi}_{K K \eta} + \left( \frac{1}{2} m_{K}^4 - 2 \frac{m_{K}^6}{m_{\pi}^2} - \frac{1}{6} m_{K}^2 m_{\pi}^2 \right) \overline{H}^{\chi}_{2K K \eta} \nonumber \\	
	& + \left( \frac{91}{108} m_{K}^4 - \frac{m_{K}^6}{m_{\pi}^2} - \frac{5}{27} m_{K}^2 m_{\pi}^2 + \frac{1}{108} m_{\pi}^4 \right) \overline{H}^{\chi}_{K K 2\eta}
\end{align}

Closed form expressions, at $\mathcal{O}(\epsilon^0)$, for the master integrals $\overline{H}^{\chi}$ appearing in ${d}_{\pi K K}$ and ${d}_{\pi \eta \eta}$ are given in Appendix \ref{SecMI}. The master integrals appearing in ${d}_{K K \eta}$ are of three mass scales, for which there exist no simple closed form expressions. For these, therefore, we take an expansion around $s=m_{\pi}^2=0$. Up to order $\mathcal{O} \left(m_{\pi}^4\right)$, we have:
\begin{align}
\label{dkkeapprox}
	(16 \pi^2)^2 \; {d}_{K K \eta} = {d}_{K K \eta}^{(-1)} (m_{\pi}^2)^{-1} + {d}_{K K \eta}^{(0)} + {d}_{K K \eta}^{(1)} (m_{\pi}^2) + {d}_{K K \eta}^{(2)} (m_{\pi}^2)^2 
\end{align}
where
\begin{align}
	{d}_{K K \eta}^{(-1)} &= \left(\frac{51}{16}+\frac{\pi ^2}{96}\right) m_{K}^6 - \frac{35}{48} m_{K}^4 m_{\pi}^2 +\left(\frac{1}{12}-\frac{\pi^2}{96}\right) m_{K}^2 m_{\pi}^4 -\frac{1}{96} m_{\pi}^6
\nonumber \\
	& \quad - \left(\frac{1}{8} m_{K}^6 + \frac{3}{32} m_{K}^4 m_{\pi}^2 - \frac{1}{32} m_{K}^2 m_{\pi}^4 \right) \log^2 \left[ \frac{4}{3} \right]
\end{align}

\begin{align}
	{d}_{K K \eta}^{(0)} =&  - \left(\frac{4235}{3456}+\frac{25 \pi ^2}{1728}\right) m_{K}^4 + \left(\frac{485}{1728}-\frac{\pi ^2}{864}\right) m_{K}^2 m_{\pi}^2 - \frac{193}{6912} m_{\pi}^4 \nonumber \\
	&  - \left(\frac{15}{32} m_{K}^4 - \frac{1}{16} m_{K}^2 m_{\pi}^2 + \frac{1}{64} m_{\pi}^4 \right) \log [\rho] + \left(\frac{1}{16} m_{K}^4 -\frac{1}{64} m_{K}^2 m_{\pi}^2 \right) \log \left[ \frac{4}{3} \right] \nonumber \\
	&  + \left(\frac{5}{72} m_{K}^4 - \frac{5}{288}  m_{K}^2 m_{\pi}^2 \right) \log^2 \left[ \frac{4}{3} \right]  + \left(\frac{1}{3} m_{K}^4 + \frac{1}{24} m_{K}^2 m_{\pi}^2 \right) F \left[ \frac{4}{3} \right]
\end{align}

\begin{align}
	{d}_{K K \eta}^{(1)} =& \left(\frac{1}{1152}+\frac{5 \pi ^2}{288}\right) m_{K}^2-\left(\frac{31}{4608}+\frac{\pi ^2}{576}\right) m_{\pi}^2 -512  \frac{m_{\pi}^4}{m_{K}^2}  + \left(\frac{17}{144} m_{K}^2 - \frac{7}{288} m_{\pi}^2 \right) \log [\rho]
\nonumber \\
	& + \left( \frac{227}{4608} m_{\pi}^2 - 512 \frac{m_{\pi}^4}{m_{K}^2} - \frac{47}{1152} m_{K}^2 \right) \log \left[\frac{4}{3}\right]  + \left(\frac{1}{96} m_{\pi}^2 - \frac{1}{24} m_{K}^2 \right) \log^2 \left[ \frac{4}{3} \right] \nonumber \\
	& - \left(\frac{7}{48} m_{K}^2 + \frac{7}{384} m_{\pi}^2 \right) F \left[\frac{4}{3}\right]
\end{align}

\begin{align}
	& \left(4 m_{K}^2 - m_{\pi}^2\right)^2 {d}_{K K \eta}^{(2)} \nonumber \\
	& \quad = - \frac{1}{\lambda ^2} \left( \frac{161}{162} m_{K}^8 - \frac{295}{324} m_{K}^6 m_{\pi}^2 + \frac{7}{12} m_{K}^4 m_{\pi}^4 + \frac{49}{55296} \frac{m_{\pi}^{10}}{m_{K}^2} - \frac{1265}{10368} m_{K}^2 m_{\pi}^6 + \frac{35}{41472} m_{\pi}^8 \right) \nonumber \\
	& \qquad + \frac{1}{\lambda^3} \bigg( \frac{5093}{243} m_{K}^{10} - \frac{1981}{162} m_{K}^8 m_{\pi}^2 + \frac{3833}{1296} m_{K}^6 m_{\pi}^4 + \frac{1}{82944} \frac{m_{\pi}^{14}}{m_{K}^4} - \frac{3431}{7776} m_{K}^4 m_{\pi}^6 \nonumber \\
	& \qquad \qquad + \frac{29}{62208} \frac{m_{\pi}^{12}}{m_{K}^2} + \frac{17}{2592} m_{K}^2 m_{\pi}^8 + \frac{103}{20736} m_{\pi}^{10} \bigg) \log \left[\frac{4}{3}\right] -\frac{\left(4 m_{K}^2-m_{\pi}^2\right)^2}{192} \log [\rho] \nonumber \\
	& \qquad  -\frac{1}{\lambda^3} \bigg( \frac{505}{36} m_{K}^{10} - \frac{63}{16} m_{K}^8 m_{\pi}^2 + \frac{5}{12} m_{K}^6 m_{\pi}^4 - \frac{13}{144} m_{K}^4 m_{\pi}^6 + \frac{1}{12288} \frac{m_{\pi}^{12}}{m_{K}^2} + \frac{3}{256} m_{K}^2 m_{\pi}^8 \nonumber \\
	& \qquad \qquad + \frac{1}{512} m_{\pi}^{10} \bigg) F \left[\frac{4}{3}\right]
\end{align}

In the above expressions, $\tau \equiv m_{\eta}^2/m_{K}^2$, $\rho \equiv m_{\pi}^2/m_{K}^2$, $\lambda \equiv - \left( 8 m_K^2 + m_{\pi}^2 \right)/3 $, and $F[x]$ is defined in Eq.(\ref{Fdef}). Note that in this expansion, divergences appear in the $m_{\pi} \rightarrow 0$ limit. The divergences from the ${d}_{K K \eta}^{(-1)}$ term cancel against the divergences in Eq.(\ref{dsunsetGMO}) and in Eq.(\ref{dlogGMO}), while those arising from the $\log[\rho]$ and $\log^2[\rho]$ in ${d}_{K K \eta}^{(0)}$ cancel against divergences in Eqs.(\ref{dlogGMO}),(\ref{dloglogGMO}) and (\ref{dpkk}). Therefore the overall $\overline{F}_{\pi}^{(6)}$ remains non-divergent in the $m^2_{\pi} \rightarrow 0$ limit.

\section{The Pion Mass to Two Loops \label{SecPionMass}}

We repeat the steps of the previous section for the pion mass. A representation for this is given in \cite{Amoros:1999dp} as:
\begin{align}
	M_{\pi}^2 = m_{\pi 0}^2 + (m_{\pi}^2)^{(4)} + (m_{\pi}^2)^{(6)}_{CT} + (m_{\pi}^2)^{(6)}_{loop} + \mathcal{O}(p^8) 
\end{align}
where $m_{\pi 0}^2 = 2 B_0 \hat{m}$ is the bare pion mass squared, and $m_{P}$ are the physical meson masses.

\begin{align}
	\frac{F_{\pi}^2}{m_{\pi}^2} (m_{\pi}^2)^{(4)} = -8 m_{\pi}^2 (L^r_{4}+L^r_{5}-2 L^r_{6}-2 L^r_{8})-16 m_{K}^2 (L^r_{4}-2 L^r_{6})+m_{\pi}^2 \left(l^r_{\pi} + \frac{1}{9}l^r_{\eta} \right)-\frac{4}{9} m_{K}^2 l^r_{\eta} 
\end{align}
\begin{align}
	- \frac{F_{\pi}^4}{16 m_{\pi}^2} & (m_{\pi}^2)^{(6)}_{CT} = 2 m_{K}^2 m_{\pi}^2 (2 C^r_{13}+C^r_{15}-2 C^r_{16}-6 C^r_{21}-2 C^r_{32})+4 m_{K}^4 (C^r_{16}-C^r_{20}-3 C^r_{21})\nonumber \\
	& + m_{\pi}^4 (2 C^r_{12}+2 C^r_{13}+C^r_{14}+C^r_{15}+3 C^r_{16}+C^r_{17}-3 C^r_{19}-5 C^r_{20}-3 C^r_{21}-2 C^r_{31}-2 C^r_{32})
\end{align}

The $( m_{\pi}^2 )^{(6)}_{loop}$ term can be subdivided into the following components:
\begin{align}
	F_{\pi}^4 (m_{\pi}^2 )^{(6)}_{loop} = {c}^{\pi}_{sunset} + c^{\pi}_{log \times log} + c^{\pi}_{log} + c^{\pi}_{log \times L_i} + c^{\pi}_{L_i} + c^{\pi}_{L_i \times L_j}
\end{align}
where

\begin{align}
	\frac{16 \pi^2}{m_{\pi}^2}  c^{\pi}_{L_i} =& \frac{2}{9} m_{\pi}^4 \left(18 L^r_{1}+37 L^r_{2}+\frac{28}{3} L^r_{3} + \frac{8}{3} L^r_{5} - 32 L^r_{7} - 16 L^r_{8} \right) \nonumber \\
	& + \frac{1}{9} m_{K}^4 \left(104 L^r_{2}+\frac{86}{3} L^r_{3} + \frac{16}{3} L^r_{5} - 64 L^r_{7} - 32 L^r_{8}\right) \nonumber \\
	& -\frac{16}{9} m_{K}^2 m_{\pi}^2 \left(L^r_{2}+\frac{1}{3}L^r_{3} + \frac{2}{3} L^r_{5} - 8 L^r_{7} - 4 L^r_{8}\right)
\end{align}

\begin{align}
	- \frac{c^{\pi}_{L_i \times L_j}}{128 m_{\pi}^2} =& (L^r_{4}-2 L^r_{6}) \left(m_{K}^4 (4 L^r_{4}+L^r_{5}-8 L^r_{6}-2 L^r_{8})+m_{K}^2 m_{\pi}^2 (4 L^r_{4}+3 L^r_{5}-8 L^r_{6}-6 L^r_{8})\right) \nonumber \\
	& +m_{\pi}^4 (L^r_{4}+L^r_{5}-2 L^r_{6}-2 L^r_{8})^2 
\end{align}

\begin{align}
	\frac{16 \pi^2}{m_{\pi}^2} c^{\pi}_{log} =& \left(\frac{1}{16} l^r_{\eta} - \frac{1199}{144} l^r_{\pi} \right) m_{\pi}^4 -\left(\frac{20}{27} l^r_{\eta} + \frac{277}{36} l^r_{K} + \frac{3}{4} l^r_{\pi} \right) m_{K}^4 \nonumber \\
	& - \left(\frac{7}{108} l^r_{\eta} + \frac{1}{3} l^r_{K} + \frac{47}{36} l^r_{\pi} \right) m_{K}^2 m_{\pi}^2   \label{clogGMO}
\end{align}

\begin{align}
	\frac{c^{\pi}_{log \times log}}{m_{\pi}^2} =&  \left(\frac{739}{324}(l^r_{\eta})^2 - \frac{43}{18} l^r_{\eta} l^r_{K} + \frac{83}{18} (l^r_{K})^2 + \frac{1}{2} l^r_{K} l^r_{\pi} - \frac{1}{4} (l^r_{\pi})^2 \right) m_{K}^4 \nonumber \\
	& + \left( \frac{3}{2} (l^r_{\pi})^2 - \frac{67}{162} (l^r_{\eta})^2 + \frac{1}{3} l^r_{\eta} l^r_{K} +\frac{20}{9} l^r_{\eta} l^r_{\pi} + \frac{2}{9} (l^r_{K})^2 - 3 l^r_{K} l^r_{\pi} \right) m_{K}^2 m_{\pi}^2 \nonumber \\
	& + \left(\frac{121}{36} (l^r_{\pi})^2 - \frac{11}{324} (l^r_{\eta})^2 - \frac{1}{3} l^r_{\eta} l^r_{\pi} \right) m_{\pi}^4 -\frac{1}{3} \frac{m_K^6}{m_{\pi}^2} (l^r_{\eta}-l^r_{K})^2
\label{cloglogGMO}
\end{align}

\begin{align}
	\frac{c^{\pi}_{log \times L_{i}}}{m_{\pi}^2} =& 16 m_{K}^2 m_{\pi}^2 \bigg( \frac{1}{9} l^r_{\eta} (16 L^r_{1}+4 L^r_{2}+4 L^r_{3}-21 L^r_{4}-8 L^r_{5}+26 L^r_{6}-24 L^r_{7}+4 L^r_{8}) \nonumber \\
	& \qquad +l^r_{K} (L^r_{4}+L^r_{5}-2 L^r_{6}-2 L^r_{8})+5 l^r_{\pi} (L^r_{4}-2 L^r_{6})\bigg) \nonumber \\
	& -8 m_{K}^4 \bigg(\frac{4}{9} l^r_{\eta} (16 L^r_{1}+4 L^r_{2}+4 L^r_{3}-18 L^r_{4}-3 L^r_{5}+20 L^r_{6}-12 L^r_{7}+2 L^r_{8}) \nonumber \\
	& \qquad + l^r_{K} (16 L^r_{1}+4 L^r_{2}+5 L^r_{3}-20 L^r_{4}-4 L^r_{5}+24 L^r_{6}+8 L^r_{8}) \bigg) \nonumber \\
	& -8 m_{\pi}^4 \bigg(\frac{1}{9} l^r_{\eta} (4 L^r_{1}+L^r_{2}+L^r_{3}-6 L^r_{4}-4 L^r_{5}+8 L^r_{6}+6 L^r_{8}) \nonumber \\
	& \qquad + l^r_{\pi} (14 L^r_{1}+8 L^r_{2}+7 L^r_{3}-18 L^r_{4}-12 L^r_{5}+32 L^r_{6}+22 L^r_{8}) \bigg) \label{clogliGMO}
\end{align}

The contribution from the sunset integrals is given by:
\begin{align}
	{c}^{\pi}_{sunset} &= \frac{1}{(16\pi^2)^2 } \bigg[ \left(1-\frac{\pi ^2}{18}\right) m_{K}^6-\left(\frac{2435}{864}+\frac{97 \pi ^2}{432}\right) m_{K}^4 m_{\pi}^2+\left(\frac{235}{144}-\frac{23 \pi ^2}{648}\right) m_{K}^2 m_{\pi}^4 \nonumber \\
	& \qquad + \left(\frac{4757}{3456}-\frac{41 \pi ^2}{1296}\right) m_{\pi}^6 \bigg]  + {c}^{\pi}_{\pi K K} + {c}^{\pi}_{\pi \eta \eta} + {c}^{\pi}_{K K \eta} \label{cSunsetGMO}
\end{align}
where
\begin{align}
	{c}^{\pi}_{\pi \eta \eta} = \left( \frac{m_{\pi}^4}{18} \right) \overline{H}^{\chi}_{\pi \eta \eta}
\label{cpee}
\end{align}
\begin{align}
	{c}^{\pi}_{\pi K K} & = \left( \frac{3}{8} m_{K}^4 + \frac{3}{4} m_{\pi}^2 m_{K}^2 - \frac{1}{8} m_{\pi}^4 \right) \overline{H}^{\chi}_{\pi K K} + \left( \frac{1}{2} m_{\pi}^6 - \frac{1}{2} m_{\pi}^2 m_{K}^4 \right) \overline{H}^{\chi}_{2\pi K K} 
\label{cpkk}
\end{align}
\begin{align}
	{c}^{\pi}_{K K \eta} =& \left( \frac{43}{36} m_{K}^2 m_{\pi}^2 -\frac{5}{8} m_{K}^4 - \frac{17}{72} m_{\pi}^4 \right) \overline{H}^{\chi}_{K K \eta} + \left( \frac{4}{3} m_{K}^6 - \frac{5}{3} m_{K}^4 m_{\pi}^2 + \frac{1}{3} m_{K}^2 m_{\pi}^4 \right) \overline{H}^{\chi}_{2K K \eta} \nonumber \\
	& + \left( \frac{2}{3} m_{K}^6 - \frac{65}{54} m_{K}^4 m_{\pi}^2 + \frac{17}{27} m_{K}^2 m_{\pi}^4 - \frac{5}{54} m_{\pi}^6 \right) \overline{H}^{\chi}_{K K 2\eta}
\label{ckkeGMO}
\end{align}

With $\rho \equiv m_{\pi}^2/m_{K}^2$ and $\tau \equiv m_{\eta}^2/m_{K}^2$, expanding ${c}^{\pi}_{K K \eta}$ about $s = m_{\pi}^2 = 0$ gives:
\begin{align}
	(16 \pi^2)^2 {c}^{\pi}_{K K \eta} = {c}_{K K \eta}^{(0)} + {c}_{K K \eta}^{(1)} (m_{\pi}^2) + {c}_{K K \eta}^{(2)} (m_{\pi}^2)^2 + \mathcal{O}((m_{\pi}^2)^3)
\end{align}
where
\begin{align}
	{c}_{K K \eta}^{(0)} &= -\left(\frac{17}{8}+\frac{\pi ^2}{144}\right) m_{K}^6 + \frac{35}{72} m_{K}^4 m_{\pi}^2 - \left( \frac{1}{18}-\frac{\pi^2}{144} \right) m_{K}^2 m_{\pi}^4 + \frac{1}{144} m_{\pi}^6 \nonumber \\
	& \qquad + \left(\frac{1}{12} m_{K}^6 +\frac{1}{16} m_{K}^4 m_{\pi}^2 - \frac{1}{48} m_{K}^2 m_{\pi}^4 \right) \log^2 \left[ \frac{4}{3} \right] 
\end{align}

\begin{align}
	{c}_{K K \eta}^{(1)} =& \left(\frac{7945}{1728} + \frac{95 \pi^2}{864}\right) m_{K}^4 - \left(\frac{751}{864} + \frac{7 \pi^2}{432} \right) m_{K}^2 m_{\pi}^2 + \frac{155}{3456} m_{\pi}^4 \nonumber \\
	& + \left(\frac{1}{96} m_{K}^2 m_{\pi}^2 - \frac{1}{24} m_{K}^4 \right) \log \left[ \frac{4}{3} \right] + \left(\frac{13}{144} m_{K}^2 m_{\pi}^2 - \frac{13}{36} m_{K}^4 \right) \log ^2 \left[ \frac{4}{3} \right] \nonumber \\
	& + \left(\frac{5}{16} m_{K}^4 -\frac{1}{24} m_{K}^2 m_{\pi}^2 + \frac{1}{96} m_{\pi}^4 \right) \log [\rho] - \left(\frac{2}{3} m_{K}^4 + \frac{1}{12} m_{K}^2 m_{\pi}^2 \right) F \left[ \frac{4}{3} \right] 
\end{align}

\begin{align}
	& ( 4 m_K^2 - m_{\pi}^2 ) {c}_{K K \eta}^{(2)} = \left(\frac{\pi ^2}{864}-\frac{109}{2304}\right) m_{\pi}^4 -\left(\frac{289}{144}+\frac{\pi ^2}{27}\right) m_{K}^4 +\left(\frac{205}{288}+\frac{\pi ^2}{216}\right) m_{K}^2 m_{\pi}^2 \nonumber \\
	& \quad  - \frac{1}{768} \frac{m_{\pi}^6}{m_{K}^2} - \frac{1}{\lambda} \left(\frac{61}{54} m_{K}^6 - \frac{23}{48} m_{K}^4 m_{\pi}^2 - \frac{1}{2304} \frac{m_{\pi}^8}{m_{K}^2} - \frac{5}{48} m_{K}^2 m_{\pi}^4 + \frac{277}{6912} m_{\pi}^6 \right) \log \left[ \frac{4}{3} \right] \nonumber \\
	& \quad - \frac{\left(4 m_{K}^2-m_{\pi}^2\right)^2}{144} \log^2 \left[\frac{4}{3}\right] -\left(\frac{20}{9} m_{K}^4 - \frac{7}{9} m_{K}^2 m_{\pi}^2 + \frac{1}{18} m_{\pi}^4 \right) \log [\rho] \nonumber \\
	& \quad + \frac{1}{\lambda} \left( \frac{13}{24} m_{K}^4 m_{\pi}^2 + \frac{11}{144} m_{K}^2 m_{\pi}^4 + \frac{5}{1728} m_{\pi}^6  -\frac{137}{27} m_{K}^6 \right) F \left[\frac{4}{3}\right]
\end{align}

The expressions of this section agree fully with those given in \cite{Kaiser:2007kf} when the eta masses here are expressed in terms of the pion and kaon masses by means of the Gell-Mann-Okubo formula. As with the expansion of the pion decay constant in $m_{\pi}^2$, here too divergences appear in the $m_{\pi}^2 \rightarrow 0$ limit. These are offset by the divergences appearing in Eqs.(\ref{clog}),(\ref{cloglog}),(\ref{csunset}) and (\ref{cpkk}) in the same limit.
In a similar way, the terms that do not vanish as $m_\pi^2\to 0$ cancel.

\section{Expansion in the Strange Quark Mass in the Isospin Limit \label{SecExp}}

As an application of the expressions presented in the preceding sections, we present their expansion in the strange quark mass, $m_s$. More specifically, for the pion decay constant, we keep the physical kaon mass constant and expand in the small quark ratio $R_q \equiv \hat{m}/m_s$ where $\hat{m} \equiv (m_u+m_d)/2 $. Our choice of such an expansion, rather than one in which we keep $m_s$ fixed and vary $\hat{m}$, is to facilitate comparison with the results given in \cite{Gasser:2007sg}.
For the pion mass we expand in $m_s$ to compare with \cite{Kaiser:2006uv}.

The isospin limit expansion of $F_{\pi}$ is:
\begin{align}
	\frac{F_{\pi}}{F_0} = 1 + d_1 \left[ \frac{M_K^2}{(4 \pi F_0)^2} \right] + d_2 \left[ \frac{M_K^2}{(4 \pi F_0)^2} \right]^2 + \mathcal{O}(m_s^3) \label{FpiSExp}
\end{align}
where
\begin{align}
	d_1 =&  8 (4 \pi )^2 L^r_4 -\frac{1}{2} \log \left[\frac{m_K^2}{\mu^2}\right] + \left\{ 8 (4 \pi )^2 (L_4^r+L_5^r)-2 \log \left[\frac{m_K^2}{\mu^2}\right] - 2 \log[2R_q] \right\} R_q \nonumber \\
	& + \left\{ 2 -8 (4 \pi )^2 (L_4^r+L_5^r)+2 \log \left[\frac{m_{K}^2}{\mu^2}\right] + 2\log[2R_q] \right\} R_q^2  + \mathcal{O}(R_q^3) \label{d1}
\end{align}	
\begin{align}
	d_2 = d_2^{\text{tree}} + d_2^{\text{loop}} \label{d2}
\end{align}
and
\begin{align}
	\frac{d_2^{\text{tree}}}{32 (4 \pi)^4} &= C_{16}^r + L_{4}^r (3 L_{4}^r + 2 L_{5}^r - 8 L_{6}^r - 4 L_{8}^r) \nonumber \\
	& + \left\{ C^r_{15}-2 C^r_{16}+6 (L^r_{4})^2 + 4 L^r_{4} L^r_{5} - 16 L^r_{4} L^r_{6} - 4 L^r_{4} L^r_{8} + 2 (L^r_{5})^2 - 8 L^r_{5} L^r_{6} - 4 L^r_{5} L^r_{8} \right\} R_q \nonumber \\	
	& + \left\{ C^r_{14}+5 C^r_{16}+C^r_{17}-3 (L^r_{4})^2 - 2 L^r_{4} L^r_{5} + 8 L^r_{4} L^r_{6} + 4 L^r_{4} L^r_{8} - 3 (L^r_{5})^2 + 4 L^r_{5} L^r_{8} \right\} R_q^2 \nonumber \\
	& + \mathcal{O}(R_q^3)
\end{align}

\begin{align}
	d_2^{\text{loop}} &= -\frac{11}{12} \log^2 \left[\frac{M_K^2}{\mu^2}\right] + \bigg( \frac{32}{9}\mathcal{D}_1^{(0)} + \frac{7}{3} - \frac{1}{3} \log \left[\frac{4}{3}\right] \bigg) \log \left[ \frac{M_K^2}{\mu^2} \right] -\frac{73}{32} + \frac{1}{3} \log \left[ \frac{4}{3} \right] \nonumber \\
	& \quad - \frac{16}{9} \bigg( \mathcal{D}_2^{(0)} - 2 \log \left[ \frac{4}{3} \right] \mathcal{D}_3^{(0)} \bigg)  + \frac{1}{3} F\left[ \frac{4}{3} \right]  \nonumber \\
	& \nonumber \\
	& + \bigg\{ \frac{5}{4} \log^2 \left[ \frac{M_K^2}{\mu^2}\right] + \left( - \frac{16}{9} \mathcal{D}_1^{(1)} + \frac{35}{12} + \frac{5}{3} \log \left[ \frac{4}{3} \right] + \frac{1}{3} \log\left[ 2R_q \right] \right) \log \left[ \frac{M_K^2}{\mu^2} \right] + \frac{157}{48}
 \nonumber \\ &
\quad + \frac{7}{6} \log \left[ \frac{4}{3} \right] 
	 - \frac{8}{9} \left( \mathcal{D}_2^{(1)} + 2 \mathcal{D}_3^{(1)} \log \left[ \frac{4}{3} \right] \right) - \frac{5}{24} F\left[ \frac{4}{3} \right]
\nonumber\\&\quad
 + \left( \frac{4}{3} \log \left[ \frac{4}{3} \right] + 16 (4 \pi)^2 (L^r_{4}-L^r_{5}+2 L^r_{8}) \right) \log \left[ 2R_q \right] \bigg\} R_q
 \nonumber \\	& \nonumber \\	
	& + \bigg\{ -\frac{41}{6} \log^2 \left[ \frac{M_K^2}{\mu^2} \right] + \left( \frac{2}{9}\mathcal{D}_1^{(2)} + \frac{101}{36} - \frac{29}{12} \log \left[ \frac{4}{3} \right] - \frac{43}{4} \log \left[ 2R_q \right] \right) \log \left[ \frac{M_K^2}{\mu^2} \right] - \frac{8455}{1536} \nonumber \\
	& \quad - \frac{61445}{18432} \log \left[ \frac{4}{3}\right] + \frac{8}{9} \left( \mathcal{D}_2^{(2)} + \mathcal{D}_3^{(2)} \log \left[ \frac{4}{3} \right] \right)+ \frac{7873}{24576} F \left[ \frac{4}{3} \right] - 5 \log^2 \left[ 2 R_q \right] \nonumber \\	
	& \quad + \left( 8 \mathcal{D}_4^{(2)} + \frac{29}{4} - 2 \log \left[ \frac{4}{3} \right] \right) \log \left[ 2 R_q \right]  \bigg\} R_q^2 + \mathcal{O}(R_q^3)
\end{align}
and

\begin{align}
	& \mathcal{D}_1^{(0)} = (4 \pi )^2 \left( 13 L^r_1+\frac{13}{4}L^r_2+\frac{61}{16}L^r_3 -\frac{51}{8} L^r_4 \right) \nonumber \\
	& \mathcal{D}_2^{(0)} = (4 \pi )^2 \left( \frac{13}{4}L^r_2+\frac{43}{48}L^r_3 \right) \nonumber \\
	& \mathcal{D}_3^{(0)} = (4 \pi )^2 \left( 4 L^r_1 + L^r_2 + L^r_3 - 3 L^r_4 \right)
\end{align}

\begin{align}
	& \mathcal{D}_1^{(1)} = (4 \pi )^2 \left( 8 L^r_{1} + 2  L^r_{2} + 2 L^r_{3} - \frac{57}{4} L^r_{4} + \frac{57}{4} L^r_{5} - 18 L^r_{8} \right) \nonumber \\
	& \mathcal{D}_2^{(1)} = (4 \pi )^2 \left( 8 L^r_{1} + \frac{4}{3} L^r_{3} - 6 L^r_{4} + 18 L^r_{5} - 36 L^r_{8} \right) \nonumber \\
	& \mathcal{D}_3^{(1)} = (4 \pi )^2 \left( 8 L^r_{1} + 2 L^r_{2} + 2 L^r_{3} - 3 L^r_{4} + 3 L^r_{5} \right)
\end{align}

\begin{align}
	& \mathcal{D}_1^{(2)} = (4 \pi )^2 \left( 584 L^r_{1} + 308 L^r_{2} + 272 L^r_{3} - 258 L^r_{4} + 234 L^r_{5} - 432 L^r_{8} \right) \nonumber \\
	& \mathcal{D}_2^{(2)} = (4 \pi )^2 \left( 5 L^r_{1} -17 L^r_{2} - \frac{11}{6} L^r_{3} - \frac{51}{2} L^r_{4} + 75 L^r_{5} - 144 L^r_{8} \right) \nonumber \\
	& \mathcal{D}_3^{(2)} = (4 \pi )^2 \left( 20 L^r_{1}+5 L^r_{2}+5 L^r_{3}-6 L^r_{4}+9 L^r_{5} \right) \nonumber \\
	& \mathcal{D}_4^{(2)} = (4 \pi )^2 \left( 14 L^r_{1}+8 L^r_{2}+7 L^r_{3}-6 L^r_{4}+5 L^r_{5}-12 L^r_{8} \right)
\end{align}

We can then connect the chiral SU(2) constant $F$ in terms of the chiral SU(3) LECs as follows:
\begin{align}
	\frac{F}{F_0} = \lim_{m_u,m_d \to 0} \frac{F_{\pi}}{F_0} =  1 + d_1 \left[ \frac{M_K^2}{(4 \pi F_0)^2} \right] + d_2 \left[ \frac{M_K^2}{(4 \pi F_0)^2} \right]^2 + \mathcal{O}(m_s^3)  \label{Fmatching}
\end{align}
where $d_1$ and $d_2$ are understood to be in the limit $m_u=m_d=0$. In this limit Eq.(\ref{FpiSExp}) agrees perfectly with the one-loop matching done in \cite{Gasser:2007sg}.

A similar expansion for the pion mass representation given in this paper is given below. In this case, we express the expansion in terms of the parameter $B_0 m_s$ rather than $M_K^2$ so as to facilitate comparison with the results of \cite{Kaiser:2006uv}.

\begin{align}
	\frac{M_{\pi}^2}{(m_u+m_d)B_0} = 1 + c_1 \left[ \frac{m_s B_0}{(4 \pi F_0)^2} \right] + c_2 \left[ \frac{m_s B_0}{(4 \pi F_0)^2} \right]^2 + \mathcal{O} (m_s^3)  \label{MpiSExp}
\end{align}
where
\begin{align}
	c_1 =& -16 (4 \pi )^2 (L_4^r - 2 L_6^r) - \frac{2}{9} \log \left[ \frac{4 B_0 m_s}{3 \mu^2} \right] \nonumber \\
	& - \left\{ 16 (4 \pi)^2 (2 L_4^r + L_5^r - 4 L_6^r -2 L_8^r ) + \frac{1}{9} + \log \left[ \frac{4}{3} \right] - \frac{8}{9} \log \left[ \frac{4 B_0 m_s}{3 \mu^2} \right] - \log \left[ 2 R_q \right] \right\} R_q \nonumber \\
	& - \left\{ \frac{1}{36} \right\} R_q^2 + \mathcal{O}(R_q^3)
\end{align}
\begin{align}	
	c_2 = c_2^{\text{tree}} + c_2^{\text{loop}}
\end{align}
and
\begin{align}
	\frac{c_2^{\text{tree}}}{64(4\pi)^4} =& -C^r_{16}+C^r_{20}+3C^r_{21} + 4 L^r_4( L^r_4 - 2 L^r_6) \nonumber \\
	& - \left\{ 2 C^r_{13}+ C^r_{15}- 2 C^r_{20}- 12 C^r_{21} - 2 C^r_{32} - 8 \left( L^r_{4} (2 L^r_{4}+L^r_{5}-4 L^r_{6}-L^r_{8})-L^r_{5} L^r_{6} \right) \right\} Q \nonumber \\
	& - \bigg\{ 2 C^r_{12}+4 C^r_{13}+C^r_{14}+2 C^r_{15}+2 C^r_{16}+C^r_{17}-3 C^r_{19}-6 C^r_{20}-12 C^r_{21}-2 C^r_{31}
 \nonumber \\	& \quad
-4 C^r_{32} - 4 \left( 2 L^r_{4}+L^r_{5} \right) \left( 2 L^r_{4}+L^r_{5}-4 L^r_{6}-2 L^r_{8} \right) \bigg\} R_q^2 + \mathcal{O} (R_q^3)
\end{align}
\begin{align}
	c_2^{\text{loop}} &= \frac{11}{12} \log^2 \left[ \frac{B_0 m_s}{\mu^2} \right] - \left( \frac{32}{9} \mathcal{C}_1^{(0)} + \frac{380}{81} - \frac{2}{9} \log \left[ \frac{4}{3} \right] \right) \log \left[ \frac{B_0 m_s}{\mu^2} \right] - \frac{38}{81} \log \left[ \frac{4}{3} \right]
  \nonumber \\	&  \quad
 + \frac{2}{9} \log^2 \left[ \frac{4}{3} \right]
 + \frac{16}{9} \left( \mathcal{C}_2^{(0)} - 2 \log \left[ \frac{4}{3} \right] \mathcal{C}_3^{(0)} \right) + \frac{73}{16} -\frac{2}{3} F \left[ \frac{4}{3} \right] \nonumber \\
	& \nonumber \\
	& + \bigg\{ \frac{97}{54} \log^2 \left[ \frac{B_0 m_s}{\mu ^2} \right] - \left( \frac{16}{9} \mathcal{C}_1^{(1)} + \frac{1549}{162} + \frac{5}{27} \log \left[ \frac{4}{3} \right] \right) \log \left[ \frac{B_0 m_s}{\mu^2} \right] - \frac{407}{324} \log \left[ \frac{4}{3} \right]
 \nonumber \\
	&  \quad
 + \frac{8}{27} \log^2 \left[ \frac{4}{3} \right]
  - \frac{8}{9} \left( \mathcal{C}_2^{(1)} + 2 \log \left[ \frac{4}{3} \right] \mathcal{C}_3^{(1)} \right) + \frac{1075}{648} - \frac{79}{144} F\left[ \frac{4}{3}\right]
\nonumber\\&\quad
 -\left(16 \mathcal{C}_4^{(1)} + \frac{4}{9} \log \left[ \frac{4}{3} \right] - \frac{5}{9}\log\left[ \frac{B_0 m_s}{\mu^2} \right] \right) \log [2R_q] \bigg\} R_q \nonumber \\
	& \nonumber \\
	& + \bigg\{ \frac{1165}{108} \log^2 \left[ \frac{B_0 m_s}{\mu^2} \right] - \left( \frac{8}{9} \mathcal{C}_1^{(2)} + \frac{6347}{324} - \frac{7}{54} \log \left[ \frac{4}{3} \right] \right) \log \left[ \frac{B_0 m_s}{\mu^2} \right] - \frac{11663}{6912} \nonumber \\
	& \quad - \frac{71117}{82944} \log \left[ \frac{4}{3} \right] - \frac{1}{54} \log^2 \left[ \frac{4}{3} \right] + \frac{4}{9} \left( \mathcal{C}_2^{(2)} - 4 \log \left[ \frac{4}{3} \right] \mathcal{C}_3^{(2)} \right) - \frac{1373}{36864} F \left[ \frac{4}{3} \right] \nonumber \\
	& \quad - \left( \frac{8}{9} \mathcal{C}_4^{(2)} + \frac{27}{2} - \frac{1}{3} \log \left[ \frac{4}{3} \right] - \frac{119}{6} \log \left[ \frac{B_0 m_s}{\mu^2} \right] \right) \log [2R_q] + \frac{17}{2} \log^2 [2R_q] \bigg\} R_q^2 + \mathcal{O} (R_q^3)
\end{align}
and
\begin{align}
	& \mathcal{C}_1^{(0)} = (4 \pi)^2 \left( 26 L^r_1 + \frac{13}{2} L^r_2 + \frac{61}{8} L^r_3 - 29 L^r_4 - \frac{13}{2} L^r_5 + 30 L^r_6 - 6 L^r_7 + 11 L^r_8 \right) \nonumber  \\
	& \mathcal{C}_2^{(0)} = (4 \pi)^2 \left( \frac{13}{2} L^r_2 + \frac{43}{24} L^r_3 + 2 L^r_4 + \frac{4}{3} L^r_5 -4 ( L^r_6 + L^r_7 + L^r_8 ) \right) \nonumber \\
	& \mathcal{C}_3^{(0)} = (4 \pi)^2 \bigg( 8 L^r_1 + 2( L^r_2 + L^r_3) - 11 L^r_4 - 2 L^r_5 + 12 L^r_6 - 6 L^r_7 + 2 L^r_8 \bigg)
\end{align}

\begin{align}
	& \mathcal{C}_1^{(1)} = (4 \pi)^2 \left( 88 L^r_{1} + 22 L^r_{2} + \frac{53}{2}L^r_{3} - 76 L^r_{4} - 26 L^r_{5} + 72 L^r_{6} + 52 L^r_{8} \right) \nonumber  \\
	& \mathcal{C}_2^{(1)} = (4 \pi)^2 \left( 88 L^r_{1} + \frac{62}{3} L^r_{3} - 86 L^r_{4} - \frac{74}{3} L^r_{5} + 80 L^r_{6} - 28 L^r_{7} + 40 L^r_{8} \right) \nonumber \\
	& \mathcal{C}_3^{(1)} = (4 \pi)^2 \left( 16 L^r_1 + 4( L^r_2 + L^r_3) - 31 L^r_4 - 8 L^r_5 + 36 L^r_6 + 16 L^r_8 \right) \nonumber \\
	& \mathcal{C}_4^{(1)} = (4 \pi)^2 \left( 3 L^r_{4} - 4 L^r_{6} \right)
\end{align}

\begin{align}
	& \mathcal{C}_1^{(2)} = (4 \pi)^2 \left( 332 L^r_{1} + 164 L^r_{2} +\frac{301}{2} L^r_{3} - 200 L^r_{4} - 78 L^r_{5} + 312 L^r_{6} + 24 L^r_{7} + 164 L^r_{8} \right) \nonumber  \\
	& \mathcal{C}_2^{(2)} = (4 \pi)^2 \left( -204 L^r_{1} + 32 L^r_{2} - \frac{151}{3} L^r_{3} + 203 L^r_{4} + \frac{100}{3} L^r_{5} - 148 L^r_{6} - 22 L^r_{7} - 74 L^r_{8} \right) \nonumber \\
	& \mathcal{C}_3^{(2)} = (4 \pi)^2 \left( 4 L^r_{1} + L^r_{2}+ L^r_{3} - 10 L^r_{4} - 3 L^r_{5} + 12 L^r_{6} + 12 L^r_{7} + 10 L^r_{8} \right)  \nonumber \\
	& \mathcal{C}_4^{(2)} = (4 \pi)^2 \left( 252 L^r_{1} + 144 L^r_{2} + 126 L^r_{3} - 108 L^r_{4} - 54 L^r_{5} + 216 L^r_{6} + 108 L^r_{8} \right) \\ \nonumber
\end{align}

From Eq.(\ref{MpiSExp}) we obtain the matching for $B$, which agrees completely with \cite{Kaiser:2006uv} in the chiral limit:
\begin{align}
	\frac{B}{B_0} = 1 + c_1 \left[ \frac{m_s B_0}{(4 \pi F_0)^2} \right] + c_2 \left[ \frac{m_s B_0}{(4 \pi F_0)^2} \right]^2 + \mathcal{O}(m_s^3)  \label{Bmatching}
\end{align}

\section{Numerical Analysis \label{SecNum}}

We present in this section a numerical analysis of the expressions given in the preceding sections, and discuss some of their implications.

\subsection{$F_\pi$}

We begin by giving a breakdown of the relative numerical contributions of the different terms constituting the $\mathcal{O}(p^6)$ term of $F_{\pi}$. As the expressions used in sections \ref{SecPionDecay} and \ref{SecPionMass} of \cite{Amoros:1999dp}
correspond to those expressed in physical meson masses, we use the physical values of the masses. The caption of Table \ref{DTable} gives the numerical input values we used. Our expressions are exact except for the approximation used for
$d^\pi_{KK\eta}$. The value calculated using the approximate expression Eq.~(\ref{dkkeapprox}) agrees with using precise numerical expressions for the sunset integrals in Eq.~(\ref{dkkeGMO}) to 8 significant digits. The parts that do not depend on the LECs are given in Table~\ref{DTable1}. The large cancellations are due to the terms that diverge for $m_\pi\rightarrow0$.
\begin{table}
\centering
\begin{tabular}{| c || c c c || c c c | c | }
\hline
& $d^\pi_{\pi KK}$ & $d^\pi_{\pi\eta\eta}$ & $d^\pi_{KK\eta}$ & ${d}^{\pi}_{\text{sunset}}$ & $d^{\pi}_{log \times log}$ & $d^{\pi}_{log}$  & Sum \\
\hline
\hline
Physical  & \multirow{2}{*}{$-93.227$} & $-0.028$ & 100.890 & $-0.381$ & 1.825 & $-8.891$ & $-7.447$ \\
GMO &  & $-0.030$ & 106.947 & $-0.482$ & 1.976 & $-8.966$ & $-7.472$ \\[1ex]
\hline
\end{tabular} \\
\caption{Numerical contributions (in units of $10^{-6}$~GeV$^{4}$) of different terms to $\left( \overline{F}_{\pi} \right)^{(6)}_{\text{loop}}$, the parts not depending on LECs.  The inputs to these were $F_\pi = F_{\pi\text{ phys}} = 0.0922$~GeV, $m_{\pi} = m_{\pi^0} = 0.1350$~GeV, $m_{K} = m_{K}^{\text{avg}} = 0.4955$~GeV, and for the physical case $m_{\eta} = 0.5479$~GeV. The renormalization scale $\mu = 0.77$ GeV.}
\label{DTable1}
\end{table}

The most recent fit of LECs with a number of different assumptions are given in Ref.~\cite{Bijnens:2014lea}.
Their main fit is called BE14 and can be found in Table 3~\cite{Bijnens:2014lea}. We show results both for the exact fit results (BE14exact) and with the two digit precision given in the reference (BE14paper). The free fit in Table 2 in \cite{Bijnens:2014lea} was done with $L_4^r$ free and a slightly different choice of $p^6$ LECs, this fit we call free-fit and finally we take the fit with the $p^6$ LECs estimated with a chiral quark model of Table 2 in \cite{Bijnens:2014lea}, labelled CQMfit. The results for the three $L_i^r$-dependent contribution, their sum and the sum including the contributions from Table~\ref{DTable1} are given in Table~\ref{DTable}.

\begin{table}
\centering
\begin{tabular}{| c || c c c | c | c |}
\hline
Fit & $d^{\pi}_{log \times L_i}$ & $d^\pi_{L_i}$ & $ d^\pi_{L_i\times L_i}$ & Sum $L_i$ & Sum \\ 
\hline
\hline
BE14exact &  7.475 & 0.064 & 0.817 &  8.356 & 0.909 \\
BE14paper &  7.456 & 0.072 & 0.841 &  8.372 & 0.925 \\
free-fit  & 12.052 & 0.391 & 2.817 & 15.260 & 7.813 \\
CQMfit    & 12.851 & 0.461 & $-0.702$ & 12.611 & 5.164 \\ [1ex]
\hline
\end{tabular} \\
\caption{Numerical contributions (in units of $10^{-6}$~GeV$^{4}$) of different terms to the $\left( \overline{F}_{\pi} \right)^{(6)}_{\text{loop}}$ of Appendix~\ref{SecPionDecayPhysical}, the part depending on the LECs. The inputs are the same as in Table~\ref{DTable1}.}
\label{DTable}
\end{table}

\begin{table}
\centering
\begin{tabular}{| c || c c c | c | c |}
\hline
Fit & $d^{\pi}_{log \times L_i}$ & $d^\pi_{L_i}$ & $ d^\pi_{L_i\times L_i}$ & Sum $L_i$ & Sum \\ 
\hline
\hline
BE14exact &  7.443 & 0.064 &   0.817 &  8.324 & 0.852 \\
BE14paper &  7.427 & 0.072 &   0.841 &  8.340 & 0.868 \\
free-fit  & 11.993 & 0.391 &   2.817 & 15.201 & 7.729 \\
CQMfit    & 12.788 & 0.461 & $-0.702$ & 12.547 & 5.075 \\ [1ex]
\hline
\end{tabular} \\
\caption{Numerical contributions (in units of $10^{-6}$~GeV$^{4}$) of different terms to the GMO simplified $\left( \overline{F}_{\pi} \right)^{(6)}_{\text{loop}}$ of Section~\ref{SecPionDecay}, the part depending on the LECs. The inputs are the same as in Table~\ref{DTable1}.}
\label{DTable2}
\end{table}

We examine the contributions calculated using the BE14exact LECs. The largest contribution arises from the $d_{log}$ term, followed by the $d_{log \times L_i}$ term. The sign of these two terms being opposite, however, reduces the overall contribution of the explicitly $\mu$-scale dependent terms to the decay constant. In absolute value terms, the bilinear chiral log terms $d_{log \times log}$ provide the next largest contribution. The bilinear $L_i$ terms are of an order of magnitude smaller. The sunsets have a relatively small contribution in absolute value terms, but due to cancellations of the other contributions, the value of $d_{\text{sunset}}$ is little over a third of the total contribution to the sum.

The sum of the contributions calculated using BE14exact (free-fit) LECs yields:
\begin{align}
	\frac{F_{\pi}}{F_0} & = 1 + \overline{F}^{(4)}_{\pi} + \left( \overline{F}^{(6)}_{\pi} \right)_{\text{loop}} + \left( \overline{F}^{(6)}_{\pi} \right)_{\text{CT}} \nonumber\\
	& = 1 + 0.2085(0.3143) + 0.0126(0.1081) + 0.0755(0.0193) \nonumber \\
	& = 1 + 0.2085(0.3143) + 0.0881(0.1274) \nonumber \\
	& = 1.2966(1.4414)
\end{align}

Using the expressions simplified using the GMO relation, we obtain:
\begin{align}
	\frac{F_{\pi}}{F_0} & = 1 + 0.2085(0.3143) + 0.0873(0.1263)
\end{align}

The value given in \cite{Bijnens:2014lea} is:
\begin{align}
	\frac{F_{\pi}}{F_0} & = 1 + 0.208(0.313) + 0.088(0.127)
\end{align}
which agrees excellently with the physical representation, and decently with the GMO simplified representation. Note that the last term has been calculated with exact $p^6$ LECs as used in \cite{Bijnens:2014lea}.

\begin{figure}
\begin{minipage}{0.45\textwidth}
\centering
\includegraphics[width=0.98\textwidth]{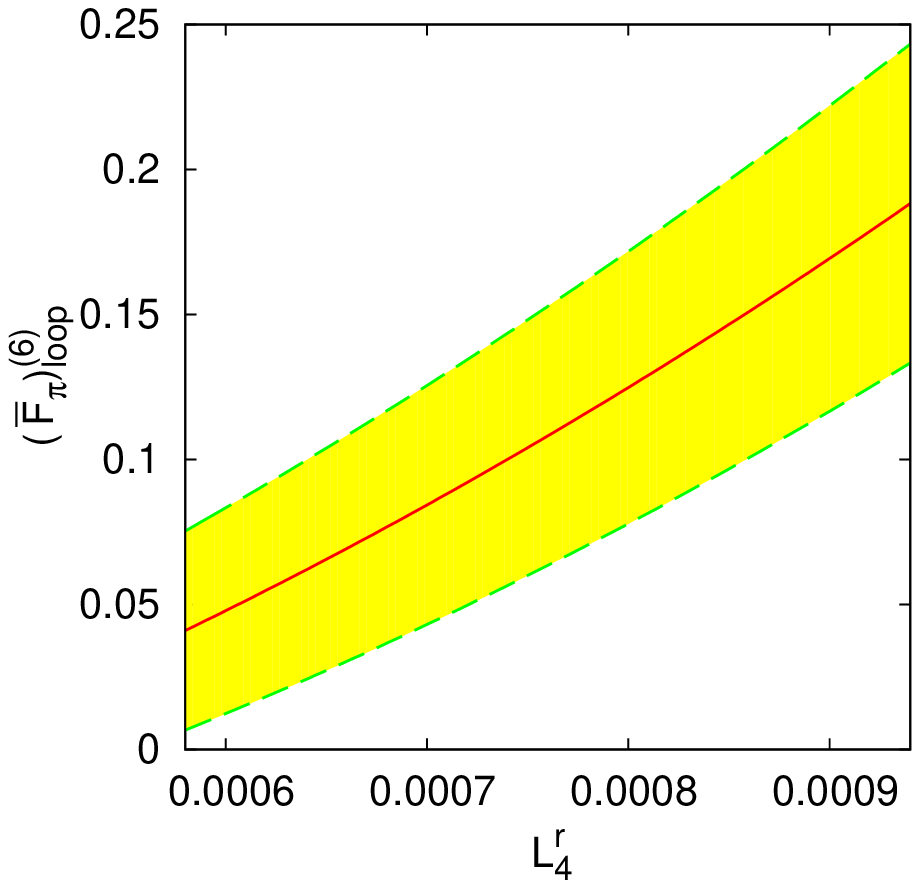} 
\caption{$L_4^r$ dependence of $(\overline{F}^{(6)}_{\pi} )_{\text{loop}} $. The full line is the value for $L_6^r=0.49 \times 10^{-3}$, while the shaded area indicates the range of possible values corresponding to the $\pm 0.25$ uncertainty of $L^r_6$ in the free fit.}
\label{FigL4Dep}
\end{minipage}
~~
\begin{minipage}{0.45\textwidth}
\includegraphics[width=0.98\textwidth]{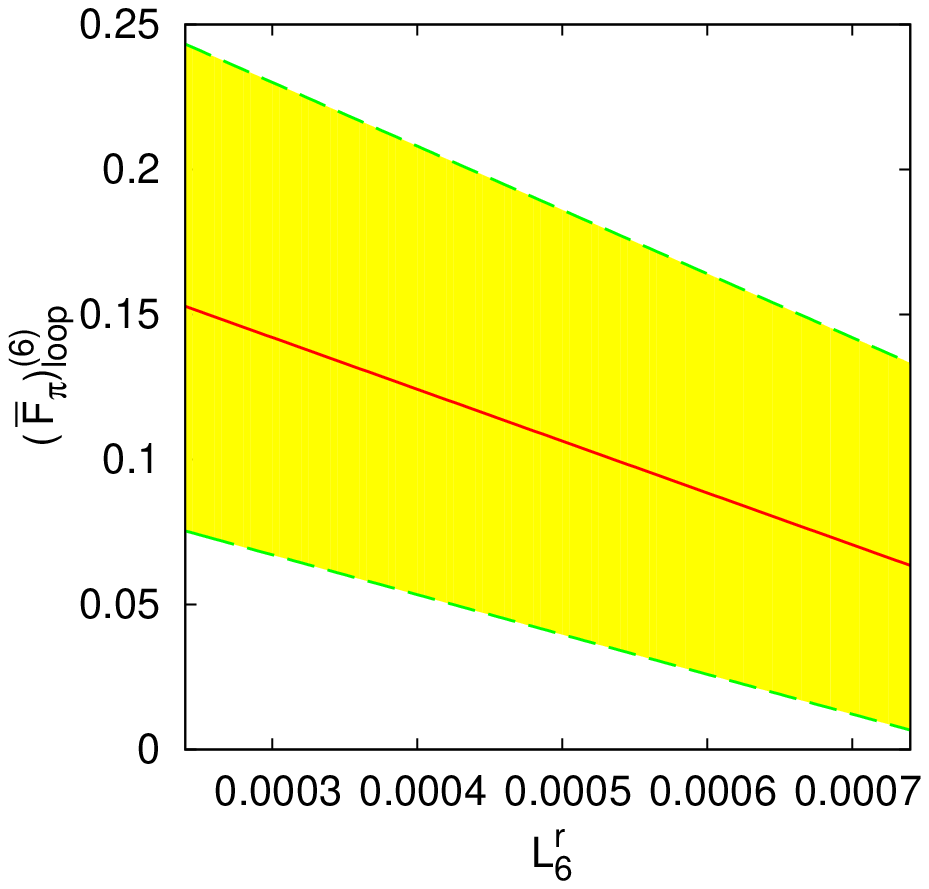} 
\caption{$L_6^r$ dependence of $(\overline{F}^{(6)}_{\pi})_{\text{loop}} $. The dashed line is the value for $L_4^r=0.76 \times 10^{-3}$, while the shaded area indicates the range of possible values corresponding to the $\pm 0.18$ uncertainty of $L^r_4$ in the free fit.}
\label{FigL6Dep}
\end{minipage}
\end{figure}

The numerical values calculated using the free-fit LECs demonstrate the sensitivity of the two-loop contribution to $F_{\pi}$ to the values of the LECs. In particular, it is to be noted that $L_4^r$ and $L_6^r$ are difficult to determine precisely, and the free fit values for these two low energy constants have relatively large uncertainties. The variation of $( \overline{F}^{(6)}_{\pi} )_{\text{loop}} $ with $L_4^r$ and $L_6^r$ over their possible range in the free fit is shown in Figures \ref{FigL4Dep} and \ref{FigL6Dep}. The trend is of a progressively smaller value of $( \overline{F}^{(6)}_{\pi} )_{\text{loop}}$ for increasing $L_6^r$ and decreasing $L_4^r$. A more thorough fit and detailed analysis of the LECs with the $F_{\pi}$ representation is planned for the future after a similar representation for the kaon and eta have been obtained.

The dependence of $F_{\pi}/F_0$ on $M_K^2$ given in Eq.(\ref{Fmatching}), with $M_K = 0.4955$ GeV and $F_0$ on the r.h.s. replaced by the physical $F_{\pi \text{ phys}}$, has the following numerical form in the chiral limit:
\begin{align}
	\frac{F}{F_0} = 1 + 0.1499(0.2562) + 0.0157(-0.0516) + ... \label{FnumChiral}
\end{align}

\begin{figure}
\centering
\includegraphics[width = 0.45\textwidth]{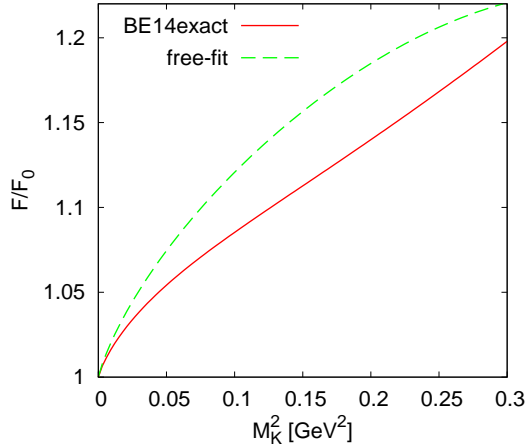}
\caption{$M^2_K$ dependence of $F/F_0$  in the chiral limit.}
\label{MsvsFpi}
\end{figure}

The first set of numbers correspond to the use of the BE14exact LECs, while
the numbers in parentheses are calculated using the free fit. Figure \ref{MsvsFpi} shows the $M_K$ dependence of $F/F_0$ using these inputs, keeping $F_0=F_\pi$ fixed on the. A significant divergence in the two sets of values is observed as $M_K^2$ increases.

The largest contribution to $F/F_0$ at $\mathcal{O}(m_s^2)$ comes from the $d_2^{\text{tree}}$ term, followed by the term proportional to $\log(B_0 m_s/\mu^2)$. In absolute terms, the pure number contribution to $d_2$ is greater than that of the $(-11/12)\log(B_0 m_s/\mu^2)$ term, but its sign being negative, the pure number serves to decrease the numerical size of $d_2$, as do all the remaining terms as well. Ignoring the terms proportional to the $L_i$ in $d_2^{\textrm{loop}}$, one gets a value of $-1.4244$ for $d_2$, in contrast to $0.4698$ when the $L_i$ proportional terms are retained. The $L_i$
therefore contribute significantly to the $\mathcal{O}(M_K^2)$ contribution to $F_{\pi}$.

The effect of the higher order terms in $R_q$ can be seen by comparing comparing Eq.(\ref{FnumChiral}) with Eq.(\ref{FnumIsospin}) below, which gives numerical values for $F_\pi/F_0$. We use a value of $R_q = \hat{m}/m_s = 1/24.4$ obtained from \cite{Leutwyler:1996qg}, the numerical value of $d_1$, Eq.(\ref{d1}), with corrections upto $\mathcal{O}(R_q^2)$ is:
\begin{align}
	d_1 &= 0.8198 (1.4009) + 0.3454 (0.3425) - 0.0108 (-0.0107) \nonumber \\
		& = 1.1544 (1.7327)
\end{align}
Similarly,
\begin{align}
	d_2^{tree} &= 2.5022 (-0.0863) - 0.3229 (-0.2641) + 0.0170 (0.0129) \nonumber \\
	& = 2.1963 (-0.3375)
\end{align}
\begin{align}
	d_2^{loop} &= -2.0324 (-1.4574) - 0.0180 (-0.1834) - 0.0729 (-0.0718) \nonumber \\
	& = -2.1233 (-1.7126)
\end{align}

Note that the $\mathcal{O}(R_q)$ contribution of $d_2^{loop}$ evaluated using the BE14exact LECs is numerically smaller than the $\mathcal{O}(R_q^2)$. Note too that the $\mathcal{O}(R_q)$ value calculated using the free fit value differs from the one calculated using BE14exact by an order of magnitude. Putting it all together we obtain up to $\mathcal{O}(R_q^2,s^2)$ the following expansion:
\begin{align}
	\frac{F_{\pi}}{F_0} = 1 + 0.2111 (0.3169) + 0.0024 (-0.0686) + \cdots \label{FnumIsospin}
\end{align}
gives a more accurate numerical representation of the effect on $F_\pi$ of integrating the strange-quark mass out. The effect of the correction due to $\hat m$ to the chiral limit is particular pronounced at $\mathcal{O}(R_q^2)$, with the value of the chiral limit number at this order given in Eq.(\ref{FnumChiral}) calculated using the BE14 fit differs from its analogous value in Eq.(\ref{FnumIsospin}) by one order of magnitude, due to cancelations between the different parts.

\subsection{$m_\pi^2$}

\begin{table}
\centering
\begin{tabular}{| c || c c c || c c c | c | }
\hline
& $c^\pi_{\pi KK}$ & $c^\pi_{\pi\eta\eta}$ & $c^\pi_{KK\eta}$ & ${c}^{\pi}_{\text{sunset}}$ & $c^{\pi}_{log \times log}$ & $c^{\pi}_{log}$  & Sum \\
\hline
\hline
Physical & \multirow{2}{*}{11.721} & 0.009 & $-10.780$ & 0.774 & 0.312 & 2.272 & 3.359 \\
GMO & & 0.010 & $-11.430$ & 0.808 & 0.284 & 2.285 & 3.376 \\[1ex]
\hline
\end{tabular} \\
\caption{Numerical contributions (in units of $10^{-7}$~GeV$^{6}$) of different terms to $\left( {m}_{\pi}^2 \right)^{(6)}_{\text{loop}}$ of Appendix~\ref{SecPionMassPhysical}, the parts not depending on LECs.  The inputs are the same as in Table~\ref{DTable1}.}
\label{CTable1}
\end{table}

\begin{table}
\centering
\begin{tabular}{| c || c c c | c | c |}
\hline
Fit & $c^{\pi}_{log \times L_i}$ & $c^\pi_{L_i}$ & $ c^\pi_{L_i\times L_i}$ & Sum $L_i$ & Sum \\ 
\hline
\hline
BE14exact &  $-1.681$ & $-0.023$ & $-0.002$ &  $-1.707$ & 1.652 \\
BE14paper &  $-1.717$ & $-0.026$ & $-0.005$ &  $-1.748$ & 1.610 \\
free-fit  &  $-1.283$ & $-0.142$ & $-0.231$ &  $-1.657$ & 1.701 \\
CQMfit    &  1.570 & $-0.168$ & $-3.844$&   $-2.442$ & 0.916
\\ [1ex]
\hline
\end{tabular} \\
\caption{Numerical contributions (in units of $10^{-7}$~GeV$^{6}$) of different terms to $\left( {m}_{\pi}^2 \right)^{(6)}_{\text{loop}}$, the part depending on the LECs. The inputs are the same as in Table~\ref{DTable1}.}
\label{CTable}
\end{table}

\begin{table}
\centering
\begin{tabular}{| c || c c c | c | c |}
\hline
Fit & $c^{\pi}_{log \times L_i}$ & $c^\pi_{L_i}$ & $ c^\pi_{L_i\times L_i}$ & Sum $L_i$ & Sum \\ 
\hline
\hline
BE14exact & $-1.730$ & 0.058 & $-0.002$ & $-1.674$ & 0.170 \\
BE14paper & $-1.765$ & 0.054 & $-0.005$ & $-1.716$ & 0.166 \\
free-fit  & $-1.319$ & $-0.080$ & $-0.232$ & $-1.631$ & 0.175 \\
CQMfit    & 1.565 & $-0.173$ & $-3.844$& $-2.452$ & 0.092 \\ [1ex]
\hline
\end{tabular} \\
\caption{Numerical contributions (in units of $10^{-7}$~GeV$^{6}$) of different terms to the GMO simplified $\left( {m}_{\pi}^2 \right)^{(6)}_{\text{loop}}$ of Section~\ref{SecPionMass}, the part depending on the LECs. The inputs are the same as in Table~\ref{DTable1}.}
\label{CTable2}
\end{table}

An analysis of the expression for the pion mass produces the numerical results given in Table \ref{CTable1} and \ref{CTable}. 
The large cancellations in the sunset contributions follow from the fact that the separate parts do not vanish in the limit $m_\pi\to 0$ but their sum does. Except for CQMfit which was not a good fit in \cite{Bijnens:2014lea}, the largest contribution comes from the pure logarithmic terms, the contribution of which, however, is cancelled to a large degree by the $\log \times L_i$ term of similar magnitude but opposite sign. The bulk of the net contribution to $(M^{(6)}_{\pi})_{\text{loop}}$ therefore comes from the sunsets diagrams and the bilinears in the chiral logs. The $c_{L_i}$ and $c_{L_i \times L_j}$ contribute very little. Using the BE14exact (free-fit) LECs, we get:
\begin{align}
	\frac{M_{\pi}^2}{m_{\pi}^2} & = 1.057(0.940) + (m_{\pi}^2)^{(4)} + (m_{\pi}^2)^{(6)}_{\text{loop}} + (m_{\pi}^2)^{(6)}_{\text{CT}} \nonumber \\
	& =	1.057(0.940) - 0.0051(0.1044) + 0.1254(0.1292) - 0.1769(-0.1732)  \nonumber \\
	& = 1.057(0.940) - 0.0051(0.1044) - 0.0515(-0.0440) \,.
\end{align}

Using the expressions simplified using the GMO relation, we get:
\begin{align}
	\frac{M_{\pi}^2}{m_{\pi}^2} & = 1.057(0.940) -0.0060(0.1035) -0.0476(-0.0407)
\end{align}

The lowest order term is determined by having the right hand side sum to 1.
This agrees well with the numerical values given in \cite{Bijnens:2014lea}.

Numerically, with $\sqrt{m_s B_0} = 0.484$ GeV, $F_0=0.0922$~GeV and BE14exact (free-fit) LECs, we have for the expansion given in Eq.(\ref{Bmatching}) in the chiral limit:
\begin{align}
	\frac{B}{B_0} = 1 + 0.0197 (0.1219) - 0.0586 (-0.1027) + ...
\end{align}

\begin{figure}
\centering
\includegraphics[width=0.5\textwidth]{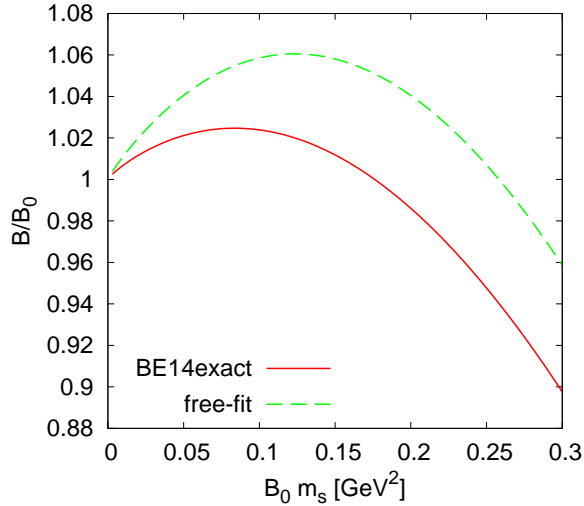}
\caption{$m_s$ dependence of $ M^2_{\pi}/m^2_{\pi}$ in the chiral limit} \label{MsvsMpi}
\end{figure}

Figure \ref{MsvsMpi} shows the $m_{s}$ dependence of $B/B_0$ for two sets of LECs, BE14exact and free-fit. Both sets of LECs produce the same general behaviour, but are different numerically. 

\section{Fitting Lattice Data \label{SecLatticeFit} }

In the equal mass case the formulas have a simple form in terms of the physical mass and decay constant. For the two-flavour case these can be found in the FLAG report \cite{Aoki:2016frl}, and for the three flavour case in \cite{Bijnens:2009qm}. Here, the only non-analytic dependences that occur are logarithms, allowing for a compact expression. Even here there are a number of different ways to express the result. In terms of the physical mass $m_\pi^2$, the physical decay constant $F_\pi$, the lowest order mass $M^2$, and the chiral limit decay constant F, the first option is:
\begin{align}
\label{eq:2flavx}
m_\pi^2 =& M^2\left\{ 1 + x \left(\frac{1}{2}\log\frac{M^2}{\mu^2}+l_M^r\right)
+ x^2 \left(\frac{17}{8}\log^2\frac{M^2}{\mu^2}+c^r_{1M}\log\frac{M^2}{\mu^2}
+c^r_{2M}\right) \right\} + \mathcal{O}\left(x^3\right)
\,,\nonumber\\
F_\pi =& F \left\{ 1 + x \left(-\log\frac{M^2}{\mu^2}+l_F^r\right)
+ x^2 \left(-\frac{5}{4}\log^2\frac{M^2}{\mu^2}+c^r_{1F}\log\frac{M^2}{\mu^2}
+c^r_{2F}\right) \right\} + \mathcal{O} \left(x^3\right)
\end{align}
Here the left-hand side is the physical observable, and the right-hand-side is expressed purely in terms of lowest order quantities. The expansion parameter here is $x=M^2/(16\pi^2 F^2)$.

An alternative is to write the lowest order on the left hand side and the physical quantities on the right hand side:
\begin{align}
\label{eq:2flavxi}
M^2 =& m_\pi^2 \left\{ 1 + \xi \left(-\frac{1}{2}\log\frac{m_\pi^2}{\mu^2}+\tilde l_M^r\right)
+ \xi^2 \left(-\frac{5}{8}\log^2\frac{m_\pi^2}{\mu^2}+\tilde c^r_{1M}\log\frac{m_\pi^2}{\mu^2}
+\tilde c^r_{2M}\right) \right\} + \mathcal{O}\left(x^3\right) 
\,,\nonumber\\
F =& F_\pi \left\{ 1 + \xi \left(\log\frac{m_\pi^2}{\mu^2}+\tilde l_F^r\right)
+ \xi^2 \left(-\frac{1}{4}\log^2\frac{m_\pi^2}{\mu^2}+
\tilde c^r_{1F}\log\frac{m_\pi^2}{\mu^2}
+\tilde c^r_{2F}\right) \right\} + \mathcal{O}\left(\xi^3\right)
\end{align}
Here the expansion is in terms of $\xi=m_\pi^2/(16\pi^2 F_\pi^2)$.

A third alternative is to have the physical quantities on the left hand side but do the expansion on the right hand side in terms of physical masses.
\begin{align}
\label{eq:2flavxip}
m_\pi^2 =& M^2 + m_\pi^2 \xi \left(\frac{1}{2}\log\frac{m_\pi^2}{\mu^2}+\hat l_M^r\right)
+ m_\pi^2 \xi^2 \left(\frac{5}{8}\log^2\frac{m_\pi^2}{\mu^2}+\hat c^r_{1M}\log\frac{m_\pi^2}{\mu^2}
+\hat c^r_{2M}\right)+\mathcal{O}\left(\xi^3\right)
\,,\nonumber\\
F_\pi =& F\left\{ 1 + \xi \left(-\log\frac{m_\pi^2}{\mu^2}+\hat l_F^r\right)
+ \xi^2 \left(\frac{5}{4}\log^2\frac{m_\pi^2}{\mu^2}+\hat c^r_{1F}\log\frac{m_\pi^2}{\mu^2}
+\hat c^r_{2F}\right) \right\} +\mathcal{O}\left(\xi^3\right)
\end{align}
There are obviously even more possibilities but these are the three that we know have been used to fit data.
The coefficients in the three options are clearly related by recursively using the expansions. The three options differ by higher orders (NNNLO).

We use a generic notation for all of the coefficients below with a $\cdot$ over the letter and $I=M,F$. The coefficients $\dot l_{I}^r,\dot c_{1I}^r$ depend on the NLO LECs while the $c_{2I}^r$ in addition depend on the NNLO LECs. The expressions (\ref{eq:2flavx}-\ref{eq:2flavxip}) are exactly $\mu$-independent when the $\mu$-dependence of the coefficients $\dot l_{I}^r,\dot c^r_{iI},\ldots$ is taken into account. The FLAG report uses a slightly different form where $l^r_I$ is traded for the scale of NLO leading logarithm $\Lambda_{3,4}$ and $c_{1I}$ for the scale of the $\log^2$ terms $\Lambda_I$ and a similar notation for the $\xi$-expansion.

A side comment is that the leading logarithms are known to higher orders \cite{Bijnens:2010xg,Bijnens:2009zi,Bijnens:2013yca}.

When different masses come into play there are clearly more ways of writing some masses as lowest order and others as physical ones, as well as the complication that the lowest order masses satisfy the Gell-Mann-Okubo (GMO) relation
allowing for having different choices for which physical masses to use. The final complication is that the non-analytic dependence from the sunset diagram is considerably more involved than just logarithms, and in fact a large aim of this program is to find faster numerical ways to handle exactly this.

In the three flavour fitting of LECs to data \cite{Amoros:2001cp,Bijnens:2011tb,Bijnens:2014lea} traditionally forms
corresponding to the third option, Eq.~(\ref{eq:2flavxip}), have been used, called ``expansion in physical masses and $F_\pi$.'' The equivalent to the $x$-expansion of Eq.~(\ref{eq:2flavx}) is usually called expansion in lowest-order quantities. Both cases were calculated in \cite{Amoros:1999dp} and can be downloaded from \cite{chpthomepage}, and are included in \textsc{CHIRON} \cite{Bijnens:2014gsa}.

In lattice calculations one has easy access to the physical masses for the charged pion and kaon while the eta mass is more difficult. On the other hand one would still like to have the expansion in terms of physical quantities since part of the higher corrections are precisely changing lowest-order masses in the loop diagrams to physical masses. For fitting lattice data we thus choose an option where one uses the physical pion decay constant and the physical charged pion and kaon masses. The eta mass in loops is then replaced by the value obtained by using the GMO relation with the physical pion and koan mass as input. These are the formulas quoted in the main text.

We can now check how many parameters are needed for the expressions for the pion mass and decay constant to NNLO. We use here the notation $m_\pi^2$ and $m_K^2$ for the physical pion and kaon masses while $m_\eta^2 = (4/3)m_K^2-(1/3)m_\pi^2$.

The GMO expressions can be written as:
\begin{align}
m_\pi^2 =& M^2 + m_\pi^2 \left\{ \frac{1}{2}\xi_\pi\lambda_\pi -\left(\frac{2}{9}\xi_K-\frac{1}{18}\xi_K\right)\lambda_\eta +\xi_K \hat L_{1M}^r + \xi_\pi \hat L_{2M}^r \right\} \nonumber\\ &
+ m_\pi^2\Bigg(  \hat K_{1M}^r \lambda_\pi^2 + \hat K_{2M}^r \lambda_\pi\lambda_K
+ \hat K_{3M}^r \lambda_\pi\lambda_\eta
+ \hat K_{4M}^r \lambda_K^2
+ \hat K_{5M}^r \lambda_K\lambda_\eta
+ \hat K_{6M}^r \lambda_\eta^2 \nonumber\\ & \hspace*{7ex}
+ \xi_K^2 F_M \left[\frac{m_\pi^2}{m_K^2}\right]
+ \hat C_{1M} \lambda_\pi+\hat C_{2M}\lambda_K+\hat C_{3M}\lambda_\eta
+ \hat C_{4M}
\Bigg)
\nonumber\\
\frac{F_\pi}{F} =& 1 + \left(-\xi_\pi\lambda_\pi-\frac{1}{2}\xi_K\lambda_K
+\xi_K \hat L_{1F}^r + \xi_\pi \hat L_{2F}^r\right)
\nonumber\\ &
+\Bigg( \hat K_{1F}^r \lambda_\pi^2
+ \hat K_{2F}^r \lambda_\pi\lambda_K
+ \hat K_{3F}^r \lambda_\pi\lambda_\eta
+ \hat K_{4F}^r \lambda_K^2
+ \hat K_{5F}^r \lambda_K\lambda_\eta
+ \hat K_{6F}^r \lambda_\eta^2
\nonumber\\ & \hspace*{7ex}
+ \xi_K^2 F_F\left[ \frac{m_\pi^2}{m_K^2} \right]
+ \hat C_{1F}\lambda_\pi+\hat C_{2F}\lambda_K+\hat C_{3F}\lambda_\eta
+ \hat C_{4F}
\Bigg)
\end{align}
where we defined the quantities $\xi_\pi=m_\pi^2/(16\pi^2 F_\pi^2)$, $\xi_K= m_K^2/(16\pi^2 F_\pi^2)$ and $\lambda_i = \log(m_i^2/\mu^2)$. The coefficients $\hat L^r_{iI}$ are a function of the NLO LECs $L_i^r$. Each of the $\hat K_{iI}^r,\hat C_{iI}^r$ has three terms proportional to $\xi_\pi^2,\xi_\pi\xi_K,\xi_K^2$ respectively.
The $\hat K_{iI}$ and $F_I$ are fully determined, the $\hat C_{iI}^r, i=1,2,3$ depend linearly on the NLO LECs
and $\hat C_{4F}$ depends up to quadratically on the NLO LECS and linearly on the NNLO LECs. There is some ambiguity in dividing the terms not depending on LECs between the various terms since $\log(m_i^2/m_K^2)=\lambda_i-\lambda_K$ for $i=\pi,\eta$.

The $F_I$ can be subdivided as \footnote{ \textcolor{red}{ A prefactor of $1/(4\pi)^4$ was included in $F_I$ in the published version of this paper, which should not be present.} }
\begin{align}
	F_I [ \rho ] &= a_{1I} + \left( a_{2I} + a_{3I} \log[\rho] + a_{4I} \log^2[\rho] \right) \rho + \left( a_{5I} + a_{6I} \log[\rho] + a_{7I} \log^2[\rho] \right) \rho^2 \nonumber \\
	& \quad + a_{8I} \log \left[ \frac{m^2_{\eta}}{m_K^2} \right] + \mathcal{O} \left( \rho^3 \right) 
\end{align}

Explicitly, the coefficients for the pion mass are given by:
\begin{align}
	& \hat{L}^r_{1M} = -16 (4 \pi )^2 (L^r_4-2L^r_6) \nonumber \\
	& \hat{L}^r_{2M} = -128 \pi ^2 (L^r_4+L^r_5-2 L^r_6-2 L^r_8)
\end{align}
\begin{align}
	& \hat{K}^r_{1M} = \frac{3}{8} \xi _{\pi} \xi_K + \frac{121}{144} \xi _{\pi }^2 \nonumber \\
	& \hat{K}^r_{2M} = -\frac{3}{4} \xi _{\pi} \xi_K \nonumber \\
	& \hat{K}^r_{3M} = \frac{5}{9} \xi _{\pi} \xi_K - \frac{1}{12} \xi _{\pi }^2 \nonumber \\
	& \hat{K}^r_{4M} = \frac{175}{144} \xi _K^2 + \frac{1}{18}\xi _{\pi } \xi _K \nonumber \\
	& \hat{K}^r_{5M} = \frac{1}{12} \xi _{\pi } \xi _K - \frac{43}{72} \xi _K^2 \nonumber \\
	& \hat{K}^r_{6M} = \frac{739}{1296} \xi _K^2 - \frac{67}{648} \xi _{\pi } \xi _K - \frac{11}{1296} \xi _{\pi }^2
\end{align}
\begin{align*}
	\hat{C}^r_{1M} =&  - \left(4(4 \pi )^2 (14 L^r_{1}+8 L^r_{2}+7 L^r_{3}-18 L^r_{4}-12 L^r_{5}+32 L^r_{6}+22 L^r_{8}) + \frac{1199}{288}\right) \xi_{\pi}^2\nonumber \\
	&  + \left(40 (4 \pi )^2 (L^r_{4}-2 L^r_{6})-\frac{47}{72}\right) \xi_{\pi } \xi_K
\end{align*}
\begin{align*}
	\hat{C}^r_{2M} =& - \left(4 (4 \pi )^2 (16 L^r_{1}+4 L^r_{2}+5 L^r_{3}-20 L^r_{4}-4 L^r_{5}+24 L^r_{6}+8 L^r_{8})+\frac{38}{9}\right) \xi_K^2 \nonumber \\
	& + \left(8 (4 \pi )^2 (L^r_{4}+L^r_{5}-2 L^r_{6}-2 L^r_{8})-\frac{1}{6}\right) \xi_{\pi } \xi_K	
\end{align*}
\begin{align*}
	\hat{C}^r_{3M} =& -\left(\frac{16}{9} (4 \pi )^2 (16 L^r_{1}+4 L^r_{2}+4 L^r_{3}-18 L^r_{4}-3 L^r_{5}+20 L^r_{6}-12 L^r_{7}+2 L^r_{8})+\frac{10}{27}\right) \xi _K^2  \nonumber \\
	& + \left(\frac{8}{9} (4 \pi )^2 (16 L^r_{1}+4 L^r_{2}+4 L^r_{3}-21 L^r_{4}-8 L^r_{5}+26 L^r_{6}-24 L^r_{7}+4 L^r_{8})-\frac{7}{216}\right) \xi_{\pi } \xi _K  \nonumber \\
	& + \left(\frac{1}{32}-\frac{4}{9} (4 \pi )^2 (4 L^r_{1}+L^r_{2}+L^r_{3}-6 L^r_{4}-4 L^r_{5}+8 L^r_{6}+6 L^r_{8})\right) \xi_{\pi}^2
\end{align*}
\begin{align}
	\hat{C}^r_{4M} &= \frac{2}{27} (4 \pi)^2 \bigg\{ (54 L^r_{1}+111 L^r_{2}+28 L^r_{3}+8 L^r_{5}-96 L^r_{7}-48 L^r_{8}) \xi_{\pi}^2  \nonumber \\
	& + (156 L^r_{2}+43 L^r_{3}+8 L^r_{5}-96 L^r_{7}-48 L^r_{8}) \xi_K^2 - 8 (3 L^r_{2}+L^r_{3}+2 L^r_{5}-24 L^r_{7}-12 L^r_{8}) \xi_\pi \xi_K \bigg\} \nonumber \\
	& -8 (8 \pi )^4 \bigg\{ (L^r_{4}-2 L^r_{6}) (4 L^r_{4}+L^r_{5}-8 L^r_{6}-2 L^r_{8}) \xi_K^2 + (L^r_{4}+L^r_{5}-2 L^r_{6}-2 L^r_{8})^2  \xi_{\pi}^2 \nonumber \\
	& - (L^r_{4}-2 L^r_{6}) (4 L^r_{4}+3 L^r_{5}-8 L^r_{6}-6 L^r_{8}) \xi_{\pi} \xi_K \bigg\} \textcolor{red}{ + 16 (4 \pi )^4 \bigg\{ (-4 C^r_{16}+4 C^r_{20}+12 C^r_{21}) \xi_K^2 } \nonumber \\
	& \textcolor{red}{  - (2 C^r_{12}+2 C^r_{13}+C^r_{14}+C^r_{15}+3 C^r_{16}+C^r_{17}-3 C^r_{19}-5 C^r_{20}-3 C^r_{21}-2 C^r_{31}-2 C^r_{32}) \xi_\pi^2 } \nonumber \\
	& \textcolor{red}{  - (4 C^r_{13}+2 C^r_{15}-4 C^r_{16}-12 C^r_{21}-4 C^r_{32}) \xi_\pi \xi_K  \bigg\} }
\label{Eq:C4M}
\end{align}

\begin{align}
	& a_{1M} = -\frac{2}{3} F\left[\frac{4}{3}\right]+\frac{73}{16}-\frac{43}{144} \log ^2\left[\frac{4}{3}\right]-\frac{1}{24} \log \left[\frac{4}{3}\right] \nonumber \\
	& a_{2M} = \frac{113}{288} F\left[\frac{4}{3}\right]-\frac{1291}{864}+\frac{1}{24} \log ^2\left[\frac{4}{3}\right]+\frac{35}{288} \log \left[\frac{4}{3}\right] \nonumber \\
	& a_{3M} = \frac{47}{72} \nonumber \\
	& a_{4M} = -\frac{3}{8} \nonumber \\
	& a_{5M} = \frac{5}{576} F\left[\frac{4}{3}\right]+\frac{8489}{6912}-\frac{1}{48} \log^2\left[\frac{4}{3}\right]-\frac{263}{2304}  \log \left[\frac{4}{3}\right] \nonumber \\
	& a_{6M} = \frac{19}{72}+\frac{1}{18} \log \left[\frac{4}{3}\right] \nonumber \\
	& a_{7M} = -\frac{31}{144} \nonumber \\
	& a_{8M} = \frac{1}{24}
\end{align}

Similiarly, for the kaon decay constant, we have:

\begin{align}
	& \hat{L}^r_{1F} = 8 (4 \pi )^2 L^r_{4} \nonumber \\
	& \hat{L}^r_{2F} = 4 (4 \pi )^2 (L^r_{4}+L^r_{5})
\end{align}
\begin{align}
	& \hat{K}^r_{1F} = \frac{41}{32} \xi _{\pi }^2 - \frac{7}{24} \xi_{\pi } \xi_K \nonumber \\
	& \hat{K}^r_{2F} = \frac{25}{12} \xi_{\pi} \xi_K \nonumber \\
	& \hat{K}^r_{3F} = 0 \nonumber \\
	& \hat{K}^r_{4F} = \frac{1}{36} \xi_{\pi} \xi_K - \frac{17}{288} \xi _K^2 \nonumber \\
	& \hat{K}^r_{5F} = \frac{1}{9} \xi_{\pi } \xi_K - \frac{55}{144} \xi _K^2 \nonumber \\
	& \hat{K}^r_{6F} = \frac{7}{288} \xi_K^2 + \frac{1}{36}\xi_{\pi} \xi_K - \frac{1}{96} \xi_{\pi }^2
\end{align}
\begin{align*}
	\hat{C}^r_{1F} =& \left(\frac{139}{144}-24 (4 \pi )^2 L^r_{4}\right) \xi _{\pi } \xi _K+\xi _{\pi }^2 \left(2 (4 \pi )^2 (14 L^r_{1}+8 L^r_{2}+7 L^r_{3}-13 L^r_{4}-10 L^r_{5})+\frac{1381}{576}\right)
\end{align*}
\begin{align*}
	\hat{C}^r_{2F} =& \left(2 (4 \pi )^2 (16 L^r_{1}+4 L^r_{2}+5 L^r_{3}-14 L^r_{4})+2\right) \xi_K^2 - \left(2 (4 \pi )^2 (3 L^r_{4}+5 L^r_{5})+\frac{1}{4}\right) \xi_{\pi} \xi_K 
\end{align*}
\begin{align*}
	\hat{C}^r_{3F} =& \left(\frac{32}{9} (4\pi)^2 (4 L^r_{1}+L^r_{2}+L^r_{3}-3 L^r_{4})+\frac{1}{3}\right) \xi_K^2 + \left(\frac{2}{9} (4\pi)^2 (4 L^r_{1}+L^r_{2}+L^r_{3}-3 L^r_{4})-\frac{11}{576} \right) \xi_{\pi}^2 \nonumber \\
	& - \left(\frac{16}{9} (4\pi)^2 (4 L^r_{1}+L^r_{2}+L^r_{3}-3 L^r_{4})+\frac{1}{144}\right) \xi_{\pi} \xi_K
\end{align*}
\begin{align} 
	\hat{C}^r_{4F} =& -\frac{1}{27} (4\pi)^2 \bigg\{ (156 L^r_{2}+43 L^r_{3}) \xi_K^2 -8 (3 L^r_{2}+L^r_{3}) \xi _{\pi } \xi _K + (54 L^r_{1}+111 L^r_{2}+28 L^r_{3}) \xi_{\pi}^2 \bigg\}  \nonumber \\
	& + (8 \pi )^4 \bigg\{ 2 \left(7 (L^r_{4})^2 + 5 L^r_{4} L^r_{5}-8 L^r_{4} L^r_{6}-4 L^r_{5} L^r_{6}\right) \xi_{\pi } \xi_K + 2 L^r_{4} (7 L^r_{4}+2 L^r_{5}-8 L^r_{6}-4 L^r_{8}) \xi_K^2 \nonumber \\
	& +\frac{1}{2} (L^r_{4}+L^r_{5}) (7 L^r_{4}+7 L^r_{5}-8 L^r_{6}-8 L^r_{8}) \xi_{\pi }^2 \bigg\} \nonumber \\
	& \textcolor{red}{ + 8 (4 \pi )^4 \bigg\{  (C^r_{14}+C^r_{15}+3 C^r_{16}+C^r_{17}) \xi_\pi^2  + 2 (C^r_{15}-2 C^r_{16}) \left(\xi _{\pi } \xi _K\right)+4 C^r_{16} \xi _K^2 \bigg\} }  
\label{Eq:C4F} 
\end{align}  \footnote{ \textcolor{red}{ The terms in red in Eqs.(\ref{Eq:C4M}) and (\ref{Eq:C4F}) were mistakenly ommitted in the published version of the paper.} }

\begin{align}
	& a_{1F} = \frac{1}{3} F\left[\frac{4}{3}\right]-\frac{73}{32}-\frac{7}{288} \log^2 \left[\frac{4}{3}\right] + \frac{1}{16} \log \left[\frac{4}{3}\right] \nonumber \\
	& a_{2F} = -\frac{5}{48} F\left[\frac{4}{3}\right]+\frac{109}{64}-\frac{1}{36} \log ^2\left[\frac{4}{3}\right]-\frac{37}{576} \log \left[\frac{4}{3}\right] \nonumber \\
	& a_{3F} = -\frac{139}{144} \nonumber \\
	& a_{4F} = \frac{7}{24} \nonumber \\
	& a_{5F} = \frac{2753}{98304} F\left[\frac{4}{3}\right]-\frac{2375}{6144}+\frac{1}{96} \log ^2\left[\frac{4}{3}\right]-\frac{533}{73728} \log \left[\frac{4}{3}\right] \nonumber \\
	& a_{6F} = \frac{47}{576} \nonumber \\
	& a_{7F} = -\frac{1}{32} \nonumber \\
	& a_{8F} = -\frac{1}{16}
\end{align}

For the equal mass case we had one free parameter at NLO for the mass
and decay constant and two each at NNLO. For the three flavour case in the
isospin limit there is a significantly larger number, two each at NLO
but, three each at NNLO not involving logarithms and 9 each for the terms
involving logarithms. The latter are clearly not independent since they at
most depend on the eight NLO LECs $L_1^r,\ldots,L_7^r$.

We defer a full study to future work when kaon and eta quantities will be included.

\section{Conclusions \label{SecConc}}

In this work, we have used the explicit representations of the two loop contribution to the pion decay constant and mass in three flavour chiral perturbation theory \cite{Amoros:1999dp} to derive (semi-)analytic expressions for them. That it is semi-analytic and not fully analytic stems from the fact that we treated the three mass configuration sunset integrals appearing in them as an expansion in the square of the external momentum and have retained only the first few terms. This semi-analytic representation is nonetheless very accurate and numerically reproduces the full result to a high degree \cite{Amoros:1999dp, Bijnens:2014gsa}.

We have used these expressions to expand $F_{\pi}$ and $M_{\pi}$ in the strange quark mass to $\mathcal{O}(m_s^2)$ and to perform the matching of two flavour low energy constants $B$ and $F$ with their three flavour counterparts in the chiral limit. The results obtained fully agree with those previously derived in \cite{Gasser:2007sg,Kaiser:2006uv,Schmid:thesis}. 

Aside from an investigation of the numerical implications of the strange quark expansion of both $F_{\pi}$ and $B_0$, we have also done a preliminary study of the dependence of $F_{\pi}$ on the low energy constants $L_4^r$ and $L_6^r$. These show trends that are possibly in contradiction with the large $N_c$ analysis of these LECs, and a more detailed study needs to be done. The breakdown of the relative numerical contributions to the pion decay constant at two loops shows that the contribution from the terms involving the $L^r_i$ and $C^r_i$, although not large, is not insignificant. Their contribution is amplified partially due
to the cancellation of other terms that have a larger absolute value. Furthermore, in the chiral limit $m_s$ expansion, the terms proportional to the low energy constants contribute greatly to the $\mathcal{O}(m_s^2)$ term. All these point to the need for a thorough study into the dependence of such quantities on the LECs for a better understanding of the chiral perturbation series.

We also present a discussion of the various ways in which NNLO results for the pion mass and decay constant may be represented, and their relative merits. We then rewrite the expressions given in this paper in a manner which allows for east fitting with data from lattice simulations.

In forthcoming work, we will present similar semi-analytic expressions for the three flavour two-loop contributions to the kaon and eta mass and decay constants, and use those results and the ones presented in this work to do a preliminary fit of lattice data to obtain new values for some low energy constants. That exercise, along with the results and analyses presented in this work, are indicative of the usefulness of such analytic representations of ChPT amplitudes and other quantities, and will hopefully encourage and facilitate the lattice community in making use of full NNLO results from ChPT.

\section*{Acknowledgements}
JB is partially supported by the Swedish Research Council grants contract numbers 621-2013-4287 and 2015-04089, and by the European Research Council (ERC) under the European Union's Horizon 2020 research and innovation programme (grant agreement No. 668679). SG thanks the authors of \cite{Gasser:2007sg, Gasser:2009hr, Gasser:2010zz} for clarifying the precise relation between their and our results, and G. Ecker, H. Leutwyler, S. Friot and M. Misiak for correspondence and discussion. BA is partly supported by the MSIL Chair of the Division of Physical and Mathematical Sciences, Indian Institute of Science.

\appendix

\section{Expressions without the use of GMO}

\subsection{Pion Mass \label{SecPionMassPhysical}}

\begin{align}
	\frac{F_{\pi}^2}{m_{\pi}^2} (m_{\pi}^2)^{(4)} = -8 m_{\pi}^2 (L^r_{4}+L^r_{5}-2 L^r_{6}-2 L^r_{8})-16 m_{K}^2 (L^r_{4}-2 L^r_{6})-\frac{1}{3} m_{\eta}^2 l^r_{\eta} + m_{\pi}^2 l^r_{\pi}
\end{align}

\begin{align}
	\frac{16 \pi^2}{m_{\pi}^2} c^{\pi}_{L_i} = m_{\pi}^4 \left(4 L^r_{1}+\frac{74}{9} L^r_{2} + \frac{56}{27} L^r_{3} \right)+\frac{1}{9} m_{K}^4 \left(104 L^r_{2}+\frac{86}{3} L^r_{3} \right) - \frac{16}{9} m_{K}^2 m_{\pi}^2 \left(L^r_{2}+\frac{1}{3}L^r_{3} \right)
\end{align}

\begin{align}
	( 16 \pi^2 ) c^{\pi}_{log} & = \left( -\frac{3}{16} m_{\eta}^4 m_{\pi}^2 + \frac{1}{4} m_{\eta}^2 m_{K}^2 m_{\pi}^2 + \frac{1}{3} m_{\eta}^2 m_{\pi}^4 - \frac{3}{4} m_{K}^4 m_{\pi}^2 - \frac{11}{6} m_{K}^2 m_{\pi}^4 - \frac{299}{36} m_{\pi}^6 \right) l_{\pi}^r \nonumber \\
	& + \left( -\frac{29}{4} m_{K}^4 m_{\pi}^2 - \frac{1}{3} m_{K}^2 m_{\pi}^4 \right) l_{K}^r + \left( \frac{3}{16} m_{\eta}^4 m_{\pi}^2 - \frac{5}{4} m_{\eta}^2 m_{K}^2 m_{\pi}^2-\frac{1}{72} m_{\eta}^2 m_{\pi}^4 \right) l_{\eta}^r \label{clog}
\end{align}

\begin{align}
	c^{\pi}_{log \times log} & = \left( \frac{121}{36} m_{\pi}^6 + \frac{3}{2} m_{\pi}^4 m_{K}^2 - \frac{1}{4} m_{\pi}^2 m_{K}^4 \right) (l_{\pi}^r)^2  + \left( \frac{1}{2} m_{\pi}^2 m_{K}^4 - 3 m_{\pi}^4 m_{K}^2 \right) l_{\pi}^r l_{K}^r
 \nonumber \\ &
  + \left( \frac{5}{3} m_{\pi}^4 m_{\eta}^2 \right) l_{\pi}^r l_{\eta}^r
	 + \left( \frac{5}{2} m_{K}^4 m_{\eta}^2 - \frac{3}{2} m_{K}^2 m_{\eta}^4 - \frac{3}{2} m_{\pi}^2 m_{K}^2 m_{\eta}^2 \right) l_{K}^r l_{\eta}^r
 \nonumber \\ &
	 + \left( \frac{1}{6} m_{\pi}^4 m_{K}^2 + \frac{19}{4} m_{\pi}^2 m_{K}^4 + \frac{1}{12} m_{\pi}^2 m_{K}^2 m_{\eta}^2 - \frac{5}{4} m_{K}^4 m_{\eta}^2 + \frac{3}{4} m_{K}^2 m_{\eta}^4 \right) (l_{K}^r)^2 \nonumber \\
	& + \left( \frac{1}{18} m_{\pi}^4 m_{\eta}^2 + \frac{25}{12}  m_{\pi}^2 m_{K}^2 m_{\eta}^2 - \frac{5}{4} m_{K}^4 m_{\eta}^2 - \frac{29}{36} m_{\pi}^2 m_{\eta}^4 + \frac{3}{4} m_{K}^2 m_{\eta}^4 \right) (l_{\eta}^r)^2
 \label{cloglog}
\end{align}

\begin{align}
	\frac{c^{\pi}_{log \times L_{i}}}{m_{\pi}^2} =& \frac{8}{9} m_{\eta}^2 m_{\pi}^2 l^r_{\eta} (12 L^r_{1}+3 L^r_{2}+3 L^r_{3}-18 L^r_{4}-8 L^r_{5}+24 L^r_{6}-48 L^r_{7}-6 L^r_{8}) \nonumber \\
	& -\frac{16}{9} m_{\eta}^2 m_{K}^2 l^r_{\eta} (24 L^r_{1}+6 L^r_{2}+6 L^r_{3}-27 L^r_{4}-4 L^r_{5}+30 L^r_{6}-24 L^r_{7}) \nonumber \\
	& -8 m_{K}^4 l^r_{K} (16 L^r_{1}+4 L^r_{2}+5 L^r_{3}-20 L^r_{4}-4 L^r_{5}+24 L^r_{6}+8 L^r_{8}) \nonumber \\
	& -8 m_{\pi}^4 l^r_{\pi} (14 L^r_{1}+8 L^r_{2}+7 L^r_{3}-18 L^r_{4}-12 L^r_{5}+32 L^r_{6}+22 L^r_{8}) \nonumber \\
	& +16 m_{K}^2 m_{\pi}^2 (l^r_{K} (L^r_{4}+L^r_{5}-2 L^r_{6}-2 L^r_{8})+5 l^r_{\pi} (L^r_{4}-2 L^r_{6}))
\end{align}

The contribution from the sunset integrals is given by:
\begin{align}
	{c}^{\pi}_{sunset} &= \frac{1}{(16\pi^2)^2 } \bigg[ \frac{3}{16} m_{\eta}^6 - \left(\frac{1}{4}+\frac{\pi^2}{16}\right) m_{\eta}^4 m_{K}^2-\frac{155}{384} m_{\eta}^4 m_{\pi}^2 + \left(\frac{9}{8}+\frac{\pi^2}{6}\right) m_{\eta}^2 m_{K}^4 \nonumber \\
	& \quad -\left(\frac{25}{32}+\frac{\pi^2}{144}\right) m_{\eta}^2 m_{K}^2 m_{\pi}^2 + \frac{25}{192} m_{\eta}^2 m_{\pi}^4 - \left(\frac{1}{2}+\frac{\pi^2}{6}\right) m_{K}^6 -\left(\frac{55}{96} + \frac{31 \pi^2}{144}\right) m_{K}^4 m_{\pi}^2 \nonumber \\
	& \quad + \left(\frac{677}{864}-\frac{5\pi^2}{162}\right) m_{K}^2 m_{\pi}^4 + \left( \frac{2543}{1728}-\frac{41\pi^2}{1296} \right) m_{\pi}^6 \bigg] + {c}^{\pi}_{\pi K K} + {c}^{\pi}_{\pi \eta \eta} + {c}^{\pi}_{K K \eta} \label{csunset}
\end{align}
where ${c}^{\pi}_{\pi \eta \eta}$ is given by Eq.(\ref{cpee}), ${c}^{\pi}_{\pi K K}$ is given by Eq.(\ref{cpkk}), and:
\begin{align}
	{c}^{\pi}_{K K \eta} & = \left(-\frac{5}{48} m_{\pi}^4 + \frac{2}{3} m_{\pi}^2 m_{K}^2 + \frac{1}{3} m_{\pi}^2 m_{\eta}^2 - \frac{5}{8} m_{K}^4 + \frac{1}{4} m_{K}^2 m_{\eta}^2 - \frac{3}{16} m_{\eta}^4 \right) \overline{H}^{\chi}_{K K \eta} \nonumber \\
	& + \left( \frac{1}{24} m_{\pi}^4 m_{K}^2 - \frac{19}{24} m_{\pi}^2 m_{K}^4 - \frac{5}{8} m_{\pi}^2 m_{K}^2 m_{\eta}^2 + \frac{5}{2} m_{K}^6 - \frac{15}{8} m_{K}^4 m_{\eta}^2 + \frac{3}{4} m_{K}^2 m_{\eta}^4 \right) \overline{H}^{\chi}_{2K K \eta}  \nonumber \\
	& + \left( \frac{7}{48} m_{\pi}^4 m_{\eta}^2 - \frac{1}{8} m_{\pi}^2 m_{K}^2 m_{\eta}^2 - \frac{1}{3} m_{\pi}^2 m_{\eta}^4 + \frac{1}{8} m_{K}^2 m_{\eta}^4 + \frac{3}{16} m_{\eta}^6 \right) \overline{H}^{\chi}_{K K 2\eta}
\end{align}

With $\rho \equiv m_{\pi}^2/m_{K}^2$ and $\tau \equiv m_{\eta}^2/m_{K}^2$, expanding ${c}^{\pi}_{K K \eta}$ about $s = m_{\pi}^2 = 0$ gives:
\begin{align}
	(16 \pi^2)^2 {c}^{\pi}_{K K \eta} = {c}_{K K \eta}^{(0)} + {c}_{K K \eta}^{(1)} (m_{\pi}^2) + {c}_{K K \eta}^{(2)} (m_{\pi}^2)^2 + \mathcal{O}((m_{\pi}^2)^3)
\end{align}
where
\begin{align}
	{c}_{K K \eta}^{(0)} &= -\frac{3}{16} m_{\eta}^6 + \left(\frac{1}{4}+\frac{\pi ^2}{16}\right) m_{\eta}^4 m_{K}^2-\left(\frac{9}{8}+\frac{\pi^2}{6}\right) m_{\eta}^2 m_{K}^4 - \left(\frac{5}{8}-\frac{5\pi^2}{48}\right) m_{K}^6 \nonumber \\
	& + \left(\frac{5}{16} m_{\eta}^2 m_{K}^4 - \frac{3}{16} m_{\eta}^4 m_{K}^2 \right) \log^2[\tau] 
\end{align}

\begin{align}
	{c}_{K K \eta}^{(1)} &= \frac{155}{384} m_{\eta}^4 + \left(\frac{353}{192}+\frac{13 \pi^2}{288}\right) m_{K}^4 + \left(\frac{49}{32}+\frac{7 \pi ^2}{144}\right) m_{\eta}^2 m_{K}^2 + \left(\frac{1}{4} m_{\eta}^2 m_{K}^2 - m_{K}^4\right) F[\tau]   \nonumber \\
	& + \left(\frac{1}{8} m_{\eta}^2 m_{K}^2 - \frac{3}{32} m_{\eta}^4 \right) \log[\tau] -\frac{13}{48} m_{\eta}^2 m_{K}^2 \log^2 [\tau] + \left(\frac{3}{32} m_{\eta}^4 - \frac{1}{8} m_{\eta}^2 m_{K}^2 +\frac{5}{16} m_{K}^4 \right) \log [\rho]
\end{align}

\begin{align}
	{c}_{K K \eta}^{(2)} &= -\left(\frac{17}{96}-\frac{\pi^2}{288}\right) m_{\eta}^2-\left(\frac{13}{48}+\frac{\pi^2}{72}\right) m_{K}^2 + \frac{1}{\lambda} \left(\frac{m_{\eta}^4}{48}-\frac{m_{K}^6}{2 m_{\eta}^2}-\frac{m_{\eta}^2 m_{K}^2}{24}-\frac{13 m_{K}^4}{24}\right) F[\tau] \nonumber \\
	& + \frac{1}{\lambda} \left(\frac{m_{\eta}^4}{6}-\frac{m_{\eta}^2 m_{K}^2}{24}-\frac{m_{K}^4}{2}\right) \log[\tau] - \frac{1}{48} m_{\eta}^2 \log^2[\tau] - \left(\frac{m_{\eta}^2}{6}+\frac{m_{K}^2}{3}\right) \log [\rho]
\end{align}

\subsection{Pion Decay Constant \label{SecPionDecayPhysical}}

\begin{align}
	(16 \pi^2) d^{\pi}_{log} &= \left( \frac{9}{32} m_{\eta}^4 - \frac{3}{8} m_{\eta}^2 m_{K}^2 - \frac{7}{48} m_{\eta}^2 m_{\pi}^2 + \frac{9}{8} m_{K}^4 + \frac{9}{4} m_{K}^2 m_{\pi}^2 + \frac{679}{144} m_{\pi}^4 \right) l_{\pi}^r \nonumber \\
	& + \left( \frac{23}{8} m_{K}^4 - \frac{1}{2} m_{K}^2 m_{\pi}^2 \right) l_{K}^r + \left( -\frac{9}{32} m_{\eta}^4 + \frac{7}{8} m_{\eta}^2 m_{K}^2 + \frac{1}{48} m_{\eta}^2 m_{\pi}^2 \right) l_{\eta}^r \label{dlog}
\end{align}

\begin{align}
	d^{\pi}_{log \times log} & = \left(\frac{15}{8} \frac{m_{K}^4 m_{\eta}^2}{m_{\pi}^2} - \frac{9}{8} \frac{m_{K}^2 m_{\eta}^4}{m_{\pi}^2} + \frac{1}{4} m_{\pi}^2 m_{\eta}^2 - \frac{17}{24}  m_{K}^2 m_{\eta}^2 + \frac{3}{8} m_{\eta}^4 \right) \left( l_{\eta}^r \right)^2
\nonumber\\&
 + \left(\frac{25}{3} m_{\pi}^2 m_{K}^2 - \frac{3}{4} m_{K}^4 \right) l_{\pi}^r l_{K}^r
 + \left(\frac{41}{8} m_{\pi}^4 - \frac{7}{6} m_{\pi}^2 m_{K}^2 + \frac{3}{8} m_{K}^4 \right) \left( l_{\pi}^r \right)^2
 \nonumber \\ 	&
 + \left(- \frac{15}{4} \frac{m_{K}^4 m_{\eta}^2}{m_{\pi}^2} + \frac{9}{4} \frac{m_{K}^2 m_{\eta}^4}{m_{\pi}^2} - \frac{7}{12} m_{K}^2 m_{\eta}^2 \right) l_{K}^r l_{\eta}^r \nonumber \\ 
	& + \left( \frac{15}{8} \frac{m_{K}^4 m_{\eta}^2}{m_{\pi}^2} - \frac{9}{8} \frac{m_{K}^2 m_{\eta}^4}{m_{\pi}^2} + \frac{1}{3} m_{\pi}^2 m_{K}^2 - \frac{5}{8} m_{K}^4 + \frac{7}{24} m_{K}^2 m_{\eta}^2 \right) \left( l_{K}^r \right)^2 \label{dloglog}
\end{align}

\begin{align}
	d^{\pi}_{log \times L_i} & = 4 m_{\pi}^2 \left(14 m_{\pi}^2 L_1^r + 8 m_{\pi}^2 L_2^r + 7 m_{\pi}^2 L_3^r - 13 m_{\pi}^2 L_4^r - 12 m_{K}^2 L_4^r - 10 m_{\pi}^2 L_5^r \right) l_{\pi}^r  \nonumber \\
&  + 4 m_{K}^2 \left(16 m_{K}^2 L_1^r + 4 m_{K}^2 L_2^r + 5 m_{K}^2 L_3^r - 3 m_{\pi}^2 L_4^r - 14 m_{K}^2 L_4^r - 5 m_{\pi}^2 L_5^r   \right) l_{K}^r \nonumber \\
&  -\frac{4}{3} m_{\eta}^2 \left( m_{\pi}^2 - 4 m_{K}^2 \right) \left( 4 L_1^r + L_2^r + L_3^r - 3 L_4^r \right) l_{\eta}^r 
\end{align}

The terms involving the sunset integrals ${d}^{\pi}_{sunset}$ is given by:
\begin{align}
	{d}^{\pi}_{sunset} &= \frac{1}{(16 \pi^2)^2} \bigg[
	- \frac{9}{32} \frac{m_{\eta}^6}{m_{\pi}^2} + \left( \frac{3}{8}+\frac{3 \pi ^2}{32}\right) \frac{ m_{\eta}^4 m_{K}^2 }{m_{\pi}^2}+\frac{193}{768} m_{\eta}^4 - \left(\frac{27}{16}+\frac{\pi^2}{4}\right) \frac{ m_{\eta}^2 m_{K}^4}{m_{\pi}^2} \nonumber \\
	& - \left(\frac{13}{64}+\frac{7 \pi ^2}{288}\right) m_{\eta}^2 m_{K}^2+\left(\frac{49}{384}+\frac{\pi^2}{216}\right) m_{\eta}^2 m_{\pi}^2 + \left(\frac{3}{4}+\frac{\pi ^2}{4}\right) \frac{m_{K}^6}{m_{\pi}^2} + \left(\frac{209}{192}+\frac{5 \pi ^2}{32}\right) m_{K}^4 \nonumber \\
	& + \left(\frac{41}{192}+\frac{\pi ^2}{36}\right) m_{K}^2 m_{\pi}^2-\left(\frac{1}{1152}+\frac{\pi ^2}{288}\right) m_{\pi}^4 \bigg] + {d}^{\pi}_{\pi K K} + {d}^{\pi}_{\pi \eta \eta} + {d}^{\pi}_{K K \eta}
\label{dsunset}
\end{align}
where ${d}^{\pi}_{\pi K K}$ is given by Eq.(\ref{dpkk}), ${d}^{\pi}_{\pi \eta \eta}$ is given by Eq.(\ref{dpee}), and:

\begin{align}
\label{dkke}
	{d}^{\pi}_{K K \eta} & = \left( \frac{1}{96} m_{\pi}^2 - \frac{1}{24} m_{K}^2 + \frac{15}{16} \frac{m_{K}^4}{m_{\pi}^2} - \frac{7}{48} m_{\eta}^2 - \frac{3}{8} \frac{m_{K}^2 m_{\eta}^2}{m_{\pi}^2} + \frac{9}{32} \frac{m_{\eta}^4}{m_{\pi}^2} \right) \overline{H}^{\chi}_{K K \eta} \nonumber \\
	& + \left( \frac{5}{48} m_{\pi}^2 m_{K}^2 - \frac{7}{48} m_{K}^4 - \frac{15}{4} \frac{m_{K}^6}{ m_{\pi}^2} + \frac{7}{16} m_{K}^2 m_{\eta}^2 +\frac{45}{16} \frac{m_{K}^4 m_{\eta}^2}{m_{\pi}^2} -\frac{9}{8} \frac{m_{K}^2 m_{\eta}^4}{m_{\pi}^2} \right) \overline{H}^{\chi}_{2K K \eta} \nonumber \\	
	& + \left( \frac{5}{96} m_{\pi}^2 m_{\eta}^2 - \frac{1}{16} m_{K}^2 m_{\eta}^2 + \frac{7}{48} m_{\eta}^4 - \frac{3}{16} \frac{m_{K}^2 m_{\eta}^4}{m_{\pi}^2} - \frac{9}{32} \frac{m_{\eta}^6}{m_{\pi}^2} \right) \overline{H}^{\chi}_{K K 2\eta}
\end{align}

This can be expressed as an expansion in $s=m_{\pi}^2$ as:
\begin{align}
	(16 \pi^2)^2 \; {d}_{K K \eta} = {d}_{K K \eta}^{(-1)} (m_{\pi}^2)^{-1} + {d}_{K K \eta}^{(0)} + {d}_{K K \eta}^{(1)} (m_{\pi}^2) + {d}_{K K \eta}^{(2)} (m_{\pi}^2)^2 + \mathcal{O} \left( (m_{\pi}^2)^3 \right)
\end{align}
where
\begin{align}
	{d}_{K K \eta}^{(-1)} &= \frac{9}{32} m_{\eta}^6 - \left(\frac{3}{8}+\frac{3 \pi^2}{32}\right) m_{\eta}^4 m_{K}^2 + \left(\frac{27}{16}+\frac{\pi^2}{4}\right) m_{\eta}^2 m_{K}^4 + \left(\frac{15}{16}-\frac{5 \pi^2}{32}\right) m_{K}^6 \nonumber \\
	& + \left(\frac{9}{32} m_{\eta}^4 m_{K}^2 - \frac{15}{32} m_{\eta}^2 m_{K}^4 \right) \log^2[\tau] 
\end{align}

\begin{align}
	{d}_{K K \eta}^{(0)} &= -\frac{193}{768} m_{\eta}^4 - \left(\frac{11}{64}-\frac{\pi ^2}{288}\right) m_{\eta}^2 m_{K}^2-\left(\frac{211}{384}+\frac{11 \pi ^2}{576}\right) m_{K}^4 + \left( \frac{1}{2} m_{K}^4 - \frac{1}{8} m_{\eta}^2 m_{K}^2 \right) F[\tau]	\nonumber \\
	& + \left( \frac{5}{96} m_{\eta}^2 m_{K}^2 \right) \log^2[\tau] + \left(\frac{9}{64} m_{\eta}^4 - \frac{3}{16} m_{\eta}^2 m_{K}^2 \right) \log[\tau]
 \nonumber \\	&
 + \left(-\frac{9}{64} m_{\eta}^4 + \frac{3}{16} m_{\eta}^2 m_{K}^2 - \frac{15}{32} m_{K}^4 \right) \log [\rho]
	\end{align}

\begin{align}
	{d}_{K K \eta}^{(1)} &= \left(\frac{19}{384}+\frac{\pi^2}{192}\right) m_{\eta}^2-\left(\frac{11}{192}-\frac{\pi^2}{96}\right) m_{K}^2 + F[\tau] \left(\frac{m_{\eta}^2}{32}-\frac{m_{K}^2}{8}\right)
 \nonumber \\	&
 + \left(\frac{7}{96} m_{\eta}^2 + \frac{1}{48} m_{K}^2 \right) \log[\rho]
 - \frac{1}{32} m_{\eta}^2 \log^2[\tau] - \frac{7}{96} \left(m_{\eta}^2\right) \log[\tau]
\end{align}

\begin{align}
	{d}_{K K \eta}^{(2)} &= \frac{1}{\lambda^2} \left( \frac{23}{576} m_{\eta}^4 - \frac{1}{4} \frac{m_{K}^6}{m_{\eta}^2} - \frac{235}{576} m_{\eta}^2 m_{K}^2 + \frac{139}{288} m_{K}^4 \right) \nonumber \\
	& + \frac{1}{\lambda^3} \left(- \frac{1}{2} \frac{m_{K}^{10}}{m_{\eta}^4} + \frac{17}{48} \frac{m_{K}^8}{m_{\eta}^2} - \frac{7}{48} m_{\eta}^2 m_{K}^4 - \frac{1}{3} m_{K}^6 \right) F[\tau] \nonumber \\
	& + \frac{1}{\lambda^3} \left(\frac{1}{192} m_{\eta}^6 - \frac{1}{32} m_{\eta}^4 m_{K}^2 - \frac{1}{2} \frac{m_{K}^8}{m_{\eta}^2} + \frac{83}{96} m_{\eta}^2 m_{K}^4 + \frac{13}{48} m_{K}^6 \right) \log [\tau] - \frac{1}{192} \log [\rho] 
\end{align}

In the above expressions, $\tau \equiv m_{\eta}^2/m_{K}^2$, $\rho \equiv m_{\pi}^2/m_{K}^2$, $\lambda \equiv m_{\eta}^2 - 4 m_K^2$, and $F[x]$ is defined in Eq.(\ref{Fdef}).

\section{Two Mass Sunset Master Integrals \label{SecMI}}

The finite parts of the master integrals appearing in the expressions for ${d}_{\pi K K}$ and ${d}_{\pi \eta \eta}$ are presented here. The chiral logarithms arising from these integrals do not appear in the expressions below, having been removed and included in the $c_{log}$, $c_{log \times log}$, $d_{log}$ or $d_{log \times log}$ term as appropriate.

\begin{align}
\overline{H}^{\chi}_{\pi K K} &= \frac{m_{K}^2}{\left(16 \pi ^2\right)^2} \bigg( 2 + \frac{\pi ^2}{6} + \frac{m_{\pi}^2}{m_{K}^2} \left(\frac{\pi ^2}{12}-\frac{1}{8}\right) - \frac{m_{\pi}^2}{2 m_{K}^2} \log^2 \left[ \frac{m_{\pi}^2}{m_{K}^2} \right] + \log \left[ \frac{m_{\pi}^2}{m_{K}^2} \right]
\nonumber \\ & \quad
 + \left(\frac{m_{K}^2}{m_{\pi}^2}+\frac{m_{\pi}^2}{m_{K}^2}-2\right) 
 \left(\text{Li}_2\left[\frac{m_{\pi}^2}{m_{K}^2}\right]+ \log \left[1-\frac{m_{\pi}^2}{m_{K}^2}\right] \log \left[\frac{m_{\pi}^2}{m_{K}^2}\right]\right) \bigg)
\end{align}

\begin{align}
\overline{H}^{\chi}_{2\pi K K} &= \frac{1}{\left(16 \pi ^2\right)^2} \bigg( \frac{\pi^2}{12} -\frac{1}{2} - \frac{1}{2}\log^2 \left[ \frac{m_{\pi}^2}{m_{K}^2} \right]
\nonumber\\&\quad
 + \left(1-\frac{m_{K}^2}{m_{\pi}^2}\right) \left( \text{Li}_2 \left[ \frac{m_{\pi}^2}{m_{K}^2} \right] + \log \left[ \frac{m_{\pi}^2}{m_{K}^2} \right] \log \left[ 1 - \frac{m_{\pi}^2}{m_{K}^2} \right] \right) \bigg)
\end{align}

The expressions for $ \overline{H}^{\chi}_{\pi \eta \eta}$ and $\overline{H}^{\chi}_{2\pi \eta \eta}$  can be obtained from the above by making the replacement $m_K \rightarrow m_{\eta}$.

\end{document}